# Persistent nonlinear phase-locking and non-monotonic energy dissipation in micromechanical resonators

Mingkang Wang[1,2], Diego J. Perez-Morelo[1,2], Daniel Lopez[3,1], and Vladimir A. Aksyuk[1]

[1]Microsystems and Nanotechnology Division, National Institute of Standards and Technology, Gaithersburg, MD 20899 USA

[2]Institute for Research in Electronics and Applied Physics, University of Maryland, College Park, MD 20742, USA

[3]Materials Research Institute, Penn State University, University Park, PA 16802

**Abstract**

Many nonlinear systems are described by eigenmodes with amplitude-dependent frequencies, interacting strongly whenever the frequencies become commensurate at internal resonances. Fast energy exchange via the resonances holds the key to rich dynamical behavior, such as time-varying relaxation rates and signatures of nonergodicity in thermal equilibrium, revealed in the recent experimental and theoretical studies of micro and nanomechanical resonators. However, a universal yet intuitive physical description for these diverse and sometimes contradictory experimental observations remains elusive. Here we experimentally reveal persistent nonlinear phase-locked states occurring at internal resonances and demonstrate that they are essential for understanding the transient dynamics of nonlinear systems with coupled eigenmodes. The measured dynamics of a fully observable micromechanical resonator system are quantitatively described by the lower frequency mode entering, maintaining, and exiting a persistent phase-locked period tripling state generated by the nonlinear driving force exerted by the higher frequency mode. This model describes the observed phase-locked coherence times, the direction and magnitude of the energy exchange, and the resulting non-monotonic mode energy evolution. Depending on the initial relative phase, the system selects distinct relaxation pathways, either entering or bypassing the locked state. The described persistent phase-locking is not limited to particular frequency fractions or types of nonlinearities and may advance nonlinear resonator systems engineering across physical domains, including photonics as well as nanomechanics.

Subject Areas: Nonlinear Dynamics, Nanophysics

# I. INTRODUCTION

Physical systems are never fully isolated from their environment, leading to unavoidable energy exchange. Understanding the origins and effects of energy dissipation has been fundamental for the development of science and engineering fields such as quantum information [1,2], cosmology [3], acoustics [4], timing [5], and sensing [6]. The linearized description of such open systems' dynamics, a bedrock of modern physics, leads to damped noninteracting eigenmodes, where damping and stochastic Langevin forces result from averaging over weak interactions with the many degrees of freedom of the external 'thermal bath' [7]. While the exact solutions for nonlinear systems are generally hard to analytically obtain, their dynamics can often be studied by treating the nonlinearities as perturbations to the linearized system, retaining the eigenmodes and eigenfrequencies, but allowing higher harmonics, amplitude-dependent eigenfrequencies and time-variable energy flows between eigenmodes. In low-loss nonlinear systems, these result in rich and complex dynamical behavior both in [8–10] and out of thermal equilibrium [11–13].

In recent years, micro and nanomechanical resonators have emerged as an ideal experimental platform to study nonlinear phenomena, due to the multiple accessible modes with high quality factors and controllable nonlinearities [9,14–19]. The individual mode nonlinearities and intermodal couplings allow the resonant frequencies to vary in time, following the changing amplitudes. Furthermore, the systems can dynamically tune in and out of the internal resonances [20–22], defined as resonances between system eigenmodes, wherein commensurate eigenfrequencies enable resonantly enhanced energy exchanges between two or more coupled modes [11]. Recent experiments involving nonlinear resonators made of carbon nanotubes (CNT), graphene membranes, and nano- and micro-mechanical structures have demonstrated interesting and unexpected dynamical behaviors both during thermalization and in thermal equilibrium. Long-time-scale energy fluctuations were observed in the thermalized state of CNT, indicative of non-ergodic behavior [10]. Multilayer graphene membrane resonators have shown abrupt transitions between two different energy relaxation regimes, with the mode-coupled state characterized by a faster relaxation rate for the observed vibrational mode [12]. In contrast, the lower-frequency vibrational mode of a nonlinear two-mode micromechanical resonator exhibited a slow, near-zero decay rate in the mode-coupled state [11]. These observations offer intriguing glimpses of non-trivial nonlinear dynamics, begging for more intuitive and universal physical descriptions of the underlying mechanisms. However, experimental progress has been hindered so far by the inability of the current experimental platforms to fully isolate and simultaneously observe all the interacting modes and to selectively prepare them in arbitrary initial excited states.

Here, we fully observe the thermalization transient dynamics of a prototypical nonlinear two-mode coupled system from an arbitrarily prepared excited state and develop a simple yet accurate quantitative model describing it. We discover that during the free-ringdown relaxation the system can either enter or bypass a persistent phase-locked state at the 1:3 internal resonance. For the given starting energies, the outcome is determined solely by the initial relative phase between the modes. The phase-locked state, arising during the free ringdown, can last up to $\approx 2.5\times$ longer than the longest thermalization time in the system and is characterized by continuous energy transfer from a higher-loss-rate mode to a lower one, decreasing the overall system dissipation and making it a nonlinear function of system energy.

The system transient dynamics can be quantitatively understood from the perspective of the lower frequency mode experiencing a time-dependent modulation force from the higher frequency mode [13], which creates stable degenerate Period Tripling States (PTS) [23–25]. While the PTS persists, the energy flows between the modes, maintaining the exact 1:3 relationship between their changing, amplitude-dependent frequencies. As a result, the lower frequency mode experiences a negative dissipation rate, exhibiting a striking non-monotonic energy evolution.

## II. ENERGY- AND PHASE-DEPENDENT RELAXATION PATHWAYS FOR COUPLED NONLINEAR RESONATORS

Fig. 1(a) depicts two isolated low-loss vibrational eigenmodes of a nonlinear micromechanical resonator used in this study, having frequencies $\omega_2/\omega_1 \approx 3$, and coupled by a cubic nonlinear term [24,25]. The lower frequency Mode 1 and the higher frequency Mode 2 have independent dissipation rates to the thermal bath, $\Gamma_1$ and $\Gamma_2$, and can exchange energy at a rate $\Gamma_{ex} \gg \Gamma_1, \Gamma_2$. Mode 1 exhibits a pronounced stiffening Duffing nonlinearity, and Mode 2 is nearly linear (see Fig. 1(b)). By independent actuation and detection at two separate frequencies, we can prepare the system at prespecified initial amplitudes (solid dots in Fig. 1(b)) and observe each mode's evolution toward thermal equilibrium, shown as the black and red arrows [see Appendix A and Supplementary Note 2].

We find that, starting from the same mode energies and depending on the initial relative phase, the system follows one of two radically distinct relaxation pathways leading to different *system energy* and oscillation frequency $\omega_1(t)$ during ringdown (Fig. 1(c),(d)). One pathway is characterized by a *phase-locked* state at the 1:3 internal resonance with a lifetime much longer than the longest thermalization time in the system (solid blue line in Fig. 1(c)). The distinctive feature of this system state is the lack of decay in the amplitude and frequency of Mode 1 supported by a constant rate of energy transfer from Mode 2. The intermodal energy transfer and the distinct dissipation rates of the modes lead to different *system* energy loss rates along the two pathways shown in Fig. 1(d).

From the perspective of Mode 1, the nonlinear coupling with Mode 2 manifests as an applied force modulated at $\omega_2 \approx 3\omega_1$, thus the equation of motion for Mode 1 can be generally written as [Supplementary Note 8]:

$$\ddot{q}_1 + \Gamma_1 \dot{q}_1 + \omega_1^2(q_1^2)\, q_1 = q_1^2 F_2(t)\cos[\phi_2(t)] \qquad (1)$$

where $q_1$ is the modal displacement, and $\omega_1(q_1^2)$ accounts for the mode's nonlinearity. Our Mode 1 is well described by a Duffing oscillator with a nonlinear coefficient $\alpha_1$: $\omega_1^2(q_1) = \omega_{1,linear}^2 + \alpha_1 q_1^2$. To obtain the 1:3 resonant coupling, the interaction with Mode 2 is described by a nonlinear force term $\propto q_1^2 q_2$, which in Eq. 1 is represented by the effective drive $F_2(t)\cos[\phi_2(t)] \propto q_2(t)$. If the nonlinear coupling can be treated as a perturbation, under the conditions discussed in Supplementary Note 8, Mode 1's effect on Mode 2 phase $\phi_2(t)$ can be neglected, and Mode 2 performs as a ringdown Duffing oscillator with gradually varying amplitude influenced by slow energy exchange with Mode 1. In this limit, the amplitude $F_2(t)$ and the frequency $\omega_2(t) = \dot{\phi}_2(t)$ vary slowly compared to the Mode 1 dynamics and the fully-coupled equations of motion can be reduced to Eq. (1) for Mode 1 subject to a period-three drive from Mode 2. As discussed below, the amplitude evolution of both modes on the long timescales of their damping rates can be quantitatively described by accounting for the energy exchange between them.

A form of Eq.1 for a Duffing oscillator with constant $F_2$ and $\omega_2$ has been a subject of recent theoretical interest [23–25] due to the existence of stable phase-locked PTS in the system, occurring when the drive force magnitude exceeds a certain nonzero threshold: $F_2 > F_{2,\text{threshold}}$ [24]. As we experimentally demonstrate below, the dynamics of our fully coupled two-mode system can be well described by the PTS model for the lower frequency Mode 1 under the influence of Mode 2 of gradually changing amplitude and near-constant frequency. Furthermore, this model may form a suitable basis for describing the dynamics of a variety of low-loss coupled nonlinear systems [see Supplementary Note 4, 7].

Such PTS for Mode 1 are schematically illustrated as solid red dots on the phase diagram shown in Fig. 1(e) and f [see Appendix B]. The sizes of the PTS basins of attraction (color-filled areas) are determined by the magnitude of the modulating force, i.e., the amplitude of Mode 2 [26]. For a substantial Mode 2 amplitude, Fig. 1(e), the PTS basins of attractions occupy much of the high-amplitude portion of the phase diagram, and the excited system is more likely to fall into the phase-locked state (solid black line) rather than bypassing it (dashed black line). Once Mode 1 is locked in a PTS, it remains there until the amplitude of decaying Mode 2 becomes so small that the PTS disappears (Fig. 1(f)). [See Supplementary Movie 1 and Note 3 for Mode 1 evolution following adiabatically decreasing Mode 2 amplitude]

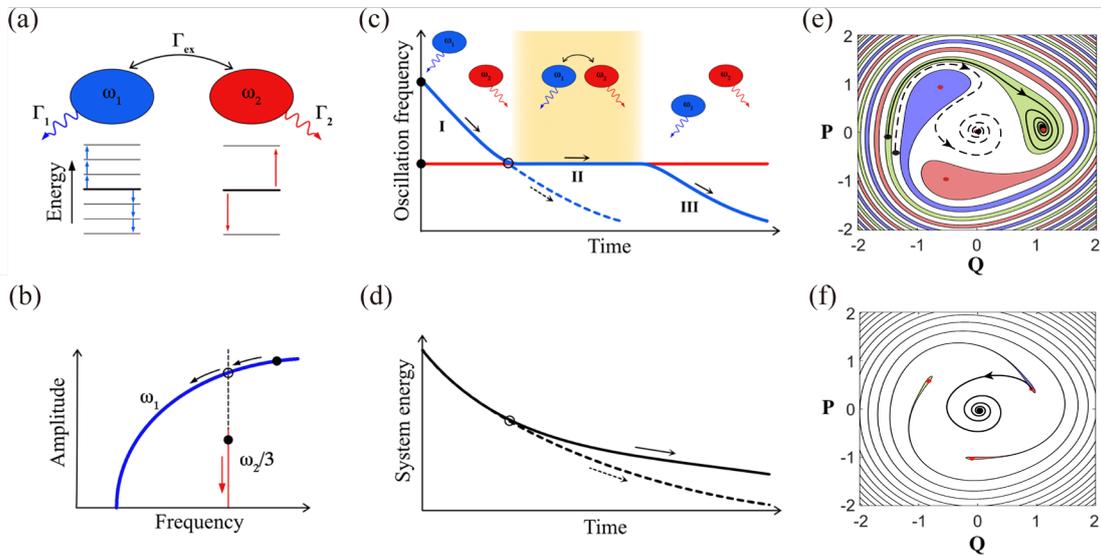

FIG. 1 Energy exchange and relaxation paths of a system consisting of two nonlinear coupled oscillators. (a) Schematic diagram showing the energy exchanged between two nonlinear coupled oscillators with commensurate frequencies (here $\omega_1:\omega_2 = 1:3$) and with the thermal bath. (b) Nonlinearity of Mode 1 allows tuning of its oscillation frequency $\omega_1$ above, at, or below the internal resonance $\omega_2/3$ as a function of Mode 1 amplitude, as shown in a schematic of the individual modes' spectral responses. (c) Oscillation frequency of Mode 1 (blue) and Mode 2 (red) during ringdown. The oscillation frequency of Mode 1 decreases during ringdown due to positive Duffing nonlinearity. At the internal resonance (black circle), Mode 1 enters (solid line) or bypass (dashed line) a phase-locked state where resonantly enhanced energy exchange with Mode 2

occurs. (d) Combined energy of the two-oscillator system. The system of coupled oscillators exhibits a different system dissipation rate (solid line) compared to that of an uncoupled system (dashed line). (e) Mode 1 dynamics can be qualitatively described by it entering (solid lines) or bypassing (dashed lines) an effective Period Tripling State (PTS), locking onto the force from Mode 2 [See Appendix B]. (f) As Mode 2 loses energy, the PTS becomes unstable and modes unlock. The simple and intuitive PTS description neglects the non-adiabatic dynamics of Mode 2. [see Supplementary Notes 7]. Compared to the experimental conditions, the schematic of the PTS separatrix uses a much smaller detuning term [see Supplementary Notes 3] for clarity of presentation. The experimental phase diagram is presented within Supplementary note 9, Figure S8.

## III. MEASUREMENT OF RELAXATION DYNAMICS NEAR INTERNAL RESONANCE

### A. Phase-locked state at internal resonance

Our experimental system is a vibrating clamped-clamped micromechanical beam that exhibits a fundamental in-plane flexural mode (Mode 1) that can couple to a higher frequency out-of-plane torsional one (Mode 2) [11,21]. After preparing the system in an initial excited state we measure the motion of both modes during subsequent free relaxation [see Supplementary Note 1], conducting two series of experiments using different initial states.

In the first series, we prepare the system initially in the phase-locked state by solely driving Mode 1 at the internal resonance [see Supplementary Note 2]. During the subsequent free ringdown, as shown in Fig. 2(a), there is a finite period of time, the "coherence time" $t_c$ (colored area in Fig. 2), where the energy of Mode 1 remains approximately constant (variation < 16 %) while Mode 2 loses energy much faster than its dissipation rate into the thermal bath. Beyond $t_c$, the modes unlock from each other and follow their individual exponential decays toward equilibrium [11,12].

The observed evolution of the amplitudes is well described by a simple model, which assumes that each mode's coupling to the thermal bath is independent and linear, and the full system energy is the sum of the energies of individual modes (i.e., time-averaged interaction energy is negligible). This condition is a good approximation for our system, though it does not generally apply to internal resonance systems. Each mode energy is assumed proportional to the square of the amplitude of the fundamental harmonic oscillation, $E_{1(2)} \propto A_{1(2)}^2$.

During $t_c$, the conservation of energy equates the extra power lost by Mode 2 to the power received by Mode 1, after accounting for their intrinsic losses. The exchange power $W$ is constant because it exactly offsets the dissipation of Mode 1, maintaining a constant amplitude on its Duffing curve at the locked frequency of $\omega_2/3$ during $t_c$. We therefore can simultaneously fit both mode amplitudes, calibrate them, and plot them on a common energy scale using only one fitting parameter $W$ [see Supplementary Note 5].

In Fig. 2(b), we plot the experimentally measured system energy as a function of time. For comparison, the dashed line corresponds to the expected energy relaxation trajectory for the system without coupling between the modes. The slower relaxation of the coupled-mode system can be explained by noting that in our system Mode 2 has a higher loss rate ($\Gamma_2/2\pi \approx$ 3.3 Hz) than Mode 1 ($\Gamma_1/2\pi \approx$ 1.5 Hz), and thus, the energy flow from a higher loss rate mode to a lower one decreases the overall system dissipation.

The temporal evolution of each mode's phase and frequency are shown in Figures 2(c)-(e). A key observation is that the modes remain phase-locked for the duration of $t_c$ (Fig. 2(c)) and, as a consequence, the relative phase $\phi_1 - \phi_2/3$ remains constant, at around 0 rad for this experiment (Fig. 2(e)), corresponding to one of the three PTS states (Fig. 2(f)). After coherence time, the modes unlock, and each mode frequency evolves toward the corresponding small-amplitude eigenfrequency, $\omega_{1,linear}/2\pi \approx 64630.6$ Hz and $\omega_{2,linear}/2\pi \approx 199881.8$ Hz.

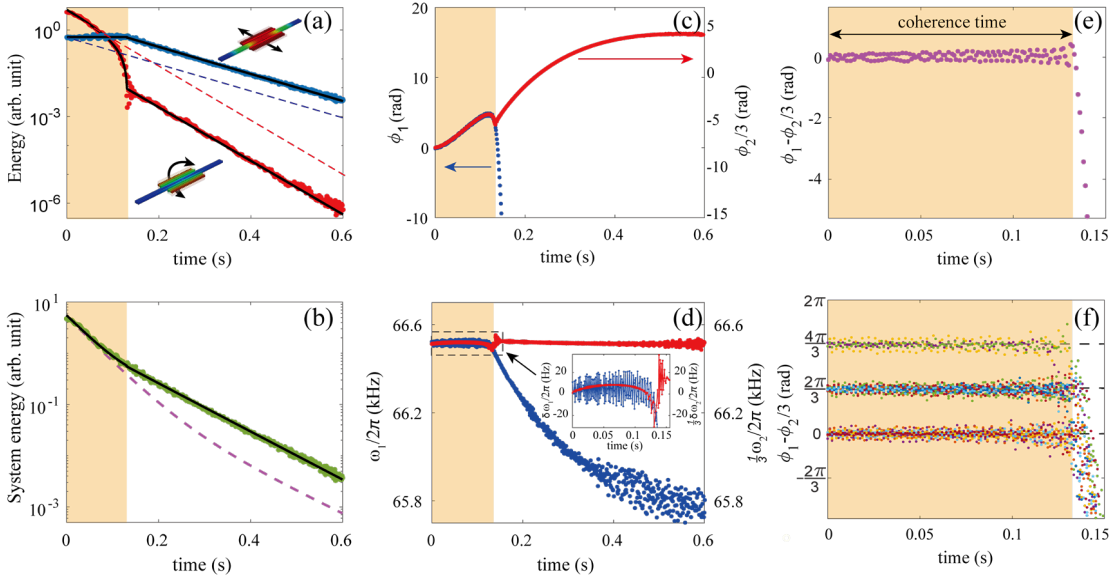

FIG. 2 Measured relaxation dynamics of Mode 1 and Mode 2 from a phase-locked state. (a) Energies $E_1$ (blue) and $E_2$ (red) for Mode 1 and 2, respectively, are plotted on a common scale, calibrated from the measured amplitudes. The black lines are fits based on energy transfer between modes. The dashed lines are hypothetic uncoupled energy decays calculated with separately measured individual mode relaxation rates. The yellow highlighted area represents the phase-locked time period. (b) Measured system energy $E_1 + E_2$ (green dots), and the model fit (black line), with the hypothetical uncoupled (purple dashed line) decay for comparison. (c) Measured phase of Mode 1 (blue dots) and Mode 2 (red dots) with reference frequency of $\omega_{2,linear}/3$ and $\omega_{2,linear}$, respectively. We define $\phi_2(t=0) = 0$ as a start phase. (d) Measured frequency. Inset shows the zoom-in of the dashed area, where $\delta\omega_1 = \omega_1 - \omega_{2,linear}/3$ and $\delta\omega_2/3 = (\omega_2 - \omega_{2,linear})/3$. (e) Measured relative phase, $\phi_1 - \phi_2/3$ is constant in the phase-locked state. (f) Repeat measurement of relative phase, $\phi_1 - \phi_2/3$ for 15 times. They show signature feature of $2\pi/3$ relative phase difference for the period-tripling states.

Fig. 2(d) shows the measured frequencies of the modes during and after coherence time. As previously mentioned, when the modes are locked, the frequencies of both Mode 1 and Mode 2 remain essentially constant (relative change $< \pm 0.05$ %). In the internal resonance Mode 1 exhibits small oscillations ($\approx 10$ Hz root mean square) around the smoothly varying $\frac{1}{3}\omega_2/2\pi \approx 66.5$ kHz. The amplitude oscillations corresponding to the frequency oscillations apparent in Fig. 2(d), and similarly Fig. 3(b), are small due to the large Duffing term [Supplementary Note 9], and cannot be resolved. Mode 2 shows gradual changes smaller than $\approx$ 10 Hz ($\approx$ 0.015 %), and a larger frequency "pull" and "rebound" of $\approx \pm$30 Hz ($\approx \pm$0.05 %) at the unlocking point. Away from the unlocking point, these minor Mode 2 deviations could be attributed to its small softening nonlinearity and off-resonant nonlinear interactions.

Since the frequency of Mode 2 remains essentially constant throughout the internal resonance, we posit that here Mode 2 can be considered an adiabatically changing external nonlinear driving force, locking Mode 1 into one of three PTS [see Supplementary Note 8]. As the system loses energy, the amplitude of Mode 2 decreases, shrinking the PTS basin of attraction, and when its amplitude drops below a fixed threshold $F_{2,\text{threshold}}$, the PTS disappears [Fig. 1(e), (f)].

We repeat the ringdown measurement multiple times and find the relative phase during locking randomly choose one of the three degenerate states of PTS. Figure 2(f) shows the relative phase of 15 groups of ringdown data with nearly the same coherent time. Their relative phase shows discrete distribution as shown in the PTS diagram shown in Fig. 1(e).

## B. Nonmonotonic energy-dependence of dissipation rate and phase-dependent relaxation pathways

One important advance in this work is that we have full experimental access with sufficient sensitivity/bandwidth to measure the dynamics of higher-frequency Mode 2, and, therefore, the relative phase, as well as the system energy, which could not be done in the previous studies [11,12,21]. Another important advance is the ability to independently control the initial conditions of both Mode 1 and Mode 2. As PTS is largely determined by the dynamics of Mode 2 ($F_2 \propto q_2$), including the relative phase as well as the amplitude, the independent initial condition control is important for understanding how the parameters of Mode 2 influence the relaxation of Mode 1, and thus the system's pathway toward equilibrium. For this purpose, in the second series of experiments we prepare the system in the unlocked initial state with starting energies above the internal resonance (black dots on the two Duffing curves in Fig. 3(a)). After this initial state is prepared by sequentially sweeping the two drive signals [See Appendix A], they are turned off simultaneously and the motion of both modes is recorded.

Since Mode 1 and Mode 2 have Duffing characteristics with opposite sign, $\omega_1$ and $\omega_2$ reach $\omega_1:\omega_2 = 1:3$ at particular amplitudes $A_{1,0}$ and $A_{2,0}$ during the ringdown, as indicated by green dots in Fig. 3(a). Before that, the modes relax independently with their frequencies changing according to their amplitude-eigenfrequency curves (Fig. 3(b)). Surprisingly, when the frequencies become commensurate, phase-locking does not always occur. We repeated the experiment up to 100 times with the same initial amplitudes but arbitrary relative phases. Once the frequencies reach the internal resonance, Mode 1 frequency either locks to Mode 2 (blue dots) or bypasses it (black dots). These results indicate that the relative phase among modes is key for entering the phase-locked state, as shown schematically in Fig. 1(e) and with experimental data in Fig. S8(c). When the modes lock, Mode 1 frequency is observed to oscillate with gradually decreasing amplitude around the Mode 2 frequency, as Mode 1 settles toward the PTS.

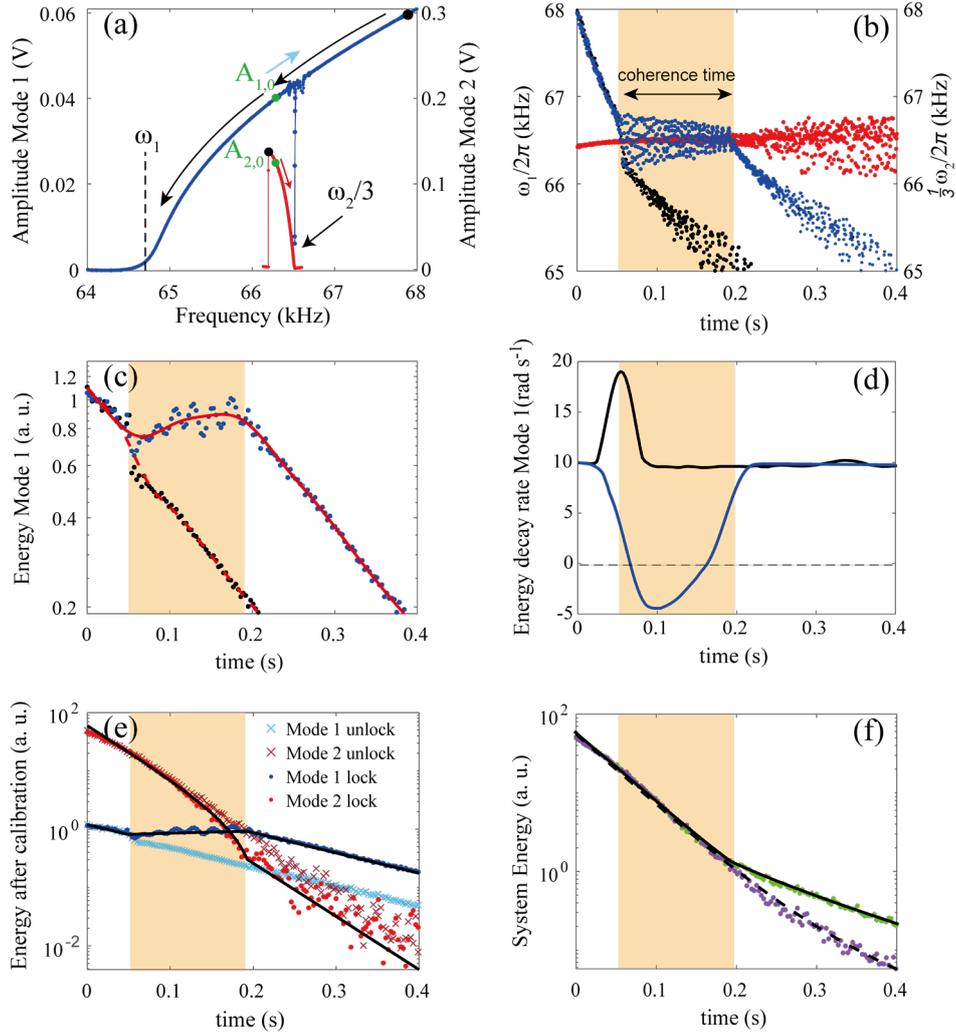

FIG. 3 Measured relaxation dynamics of Mode 1 and Mode 2 from an initially uncoupled high energy state. (a) The modes are prepared at the initial amplitudes marked by black dots by independently driving them along the shown trajectories. With the drives off, following an initial free decay, the modes reach internal resonance at the amplitudes labeled by green dots, and either enter or bypass the phase-locked state, depending on their relative phase. The arrows show the Mode 2 amplitude-frequency trajectory (red arrow) and the Mode 1 trajectory for the unlocked (black arrow) and the locked (blue arrow) states. Both modes' amplitude-frequency trajectories are confined to their respective individual nonlinear eigenfrequency-amplitude relationships. The narrow dip on the amplitude-frequency curve of the driven Mode 1 corresponds to crossing the 1:3 internal resonance at $\omega_2/3$, while the drive to Mode 2 is off. (b) Oscillation frequency of locked Mode 1 (blue dots) and Mode 2 (red dots), with comparison to that of the unlocked Mode 1 (black dots). (c) Energy of Mode 1. Blue and black dots correspond to the locked and unlocked cases. The amplitude oscillations corresponding to the frequency oscillations apparent in (b) are small due to the large Duffing term [Supplementary Note 9], and are not visible in the energy data downsampled to reduce noise. The dashed (solid) line is a 0.03 s moving average of the unlocked (locked) data. (d) Energy decay rate of Mode 1 extracted from the averaged locked (blue) and unlocked (black) data in c. The increasing energy of Mode 1 is supplied by Mode 2. (e) Two distinct energy relaxation trajectories with different final energies from the same initial energies for Mode 1 and Mode 2. (f) System

energy pathways. Black lines are obtained from the calibration based on energy transfer between modes [see Supplementary Note 5].

Fig. 3(c) shows the Mode 1 energy for the phase-locked and unlocked cases. When the modes lock, the energy can flow between them, while the time-averaged interaction energy remains low compared to the individual mode energies – no abrupt frequency shifts are observed at the locking and unlocking times and the modes continue to follow their individual amplitude-frequency relationships. In the locked state, the net system energy continues to decrease through dissipation, while the frequencies of both modes evolve together in the direction that lowers the net energy, following their frequency-amplitude relationships. Here, $\omega_1 = \omega_2/3$ increase together and Mode 2 amplitude drops, while Mode 1 amplitude increases with time [Fig. 3(a) blue arrow].

Separate from the oscillations observed in Mode 1 frequency as it settles toward the PTS, small slow variations in the energy of Mode 1 are observed, possibly arising from the non-adiabatic evolution of the pseudopotential. The resulting dynamic energy exchange between the modes is not described by the simple PTS model, requiring further investigation. For the case where locking is bypassed (black dots), a sharp loss rate increase is visible at the bypassing of internal resonance, consistent with the PTS model [see Supplementary Note 3]. Qualitatively, for a certain range of the relative phase, energy transfers away from Mode 1 to Mode 2, therefore briefly accelerating the Mode 1 decay.

The measured time-averaged energy decay rate of Mode 1 is shown in Fig. 3(d) for the locked (blue) and unlocked state (black). As it is evident from the data, the coupling among modes qualitatively changes the effective energy decay rate of Mode 1: it exhibits a striking non-monotonic energy evolution with a negative dissipation rate. When the modes do not lock, the energy loss of Mode 1 remains positive during the entire relaxation time, with a transient peak observed when the frequencies cross the internal resonance. This transient rapid energy loss observed during bypassing is distinct from the sustained faster energy loss observed at higher amplitudes in Ref. 12. It indicates that the energy gain and loss shown in here and Ref. [12], respectively, are not due to their possibly different initial conditions.

The energies of both modes are displayed in Fig. 3(e) and their sum in Fig. 3(f). The data clearly demonstrates that there are two distinct energy relaxation pathways for the system and that they are determined solely by the initial relative phase between the two modes [Supplementary Note 9]. Furthermore, because of the ability of the system to dynamically shift the energy between multiple modes with different modal dissipation rates, the system energy dissipation becomes a nonlinear function of the net system energy for as long as the modes are locked.

### C. Tunable coherence time and phase-lock probability

With the added flexibility of controlling the initial conditions of the two modes independently, we observe different coherence times. We discover that $t_c$ can be controlled ~~solely~~ by changing the amplitude of Mode 2 at locking, $A_{2,0}$, and is nearly independent of the other state variables [see Supplementary Note 5]. Fig. 4(a) shows the measured $t_c$ normalized by the longest thermalization time constant $1/\Gamma_1$ in the system, for different $A_{2,0}$. The blue squares correspond to the coherence times measured when Mode 1 is prepared initially at the internal resonance (data shown in Fig. 2), and the green diamonds show the measured coherence times when Mode 1 and 2 decay from an initially unlocked state (data shown in Fig. 3). Among all our experiments, which differ by their initial states, $t_c$ is fully determined by $A_{2,0}$, independent of any other variables describing the state of our system. Indeed, the $t_c$ is given by the time needed for $A_2$ to decay form $A_{2,0}$ to a threshold amplitude $A_{2,threshold}$, a minimum amplitude required to maintain PTS, which is a property of the system independent of its state. As $A_{2,0}$ is controllable, $A_{2,threshold}$ is directly measured (black dashed line in Fig. 4(a)), and the energy dissipation of Mode 2 is fully captured based on energy transfer between modes (fits in Fig. 2, 3 using Eq. S7 ), $t_c$ can be obtained via calculating the ringdown time of Mode 2 from $A_{2,0}$ to $A_{2,threshold}$, Eq. S8 [see Supplementary Note 5]. The calculated coherence time is shown as the black line in Fig. 4(a) without using any adjustable parameters.

While the tunable $t_c$ and non-monotonic energy-dependent dissipation rates offer a novel strategy for dissipation engineering, the probabilistic nature of reaching a phase-lock state during thermalization from a random relative phase must be taken into consideration. When the modes are relaxing from an unlocked state, the initial relative phase of the modes can be set independently from other parameters. Fig. 4(b) shows the statistical results for 100 energy decay measurements under random initial relative phase conditions for each $A_{2,0}$. The data indicate that a minimum $A_{2,0} = A_{2,threshold}$ is required to reach the phase-locking condition and that the likelihood of locking increases with $A_{2,0}$ approaching unity for large amplitudes. These observations are in agreement with the PTS [24,26] based model, where the basins of attraction for the stable states become larger with increasing $A_{2,0} \propto F_2$. Figures 4(c) and 4(d) are schematic of the PTS basins of attraction for large and small $A_{2,0}$ respectively. The purple dashed circle depicts the Mode 1 states with a given amplitude and random initial relative phases. For larger $A_{2,0}$, the majority of the points on the ring lie in the color-filled PTS basins of attraction and thus, the chance for Mode 1 to fall into a PTS is significantly larger.

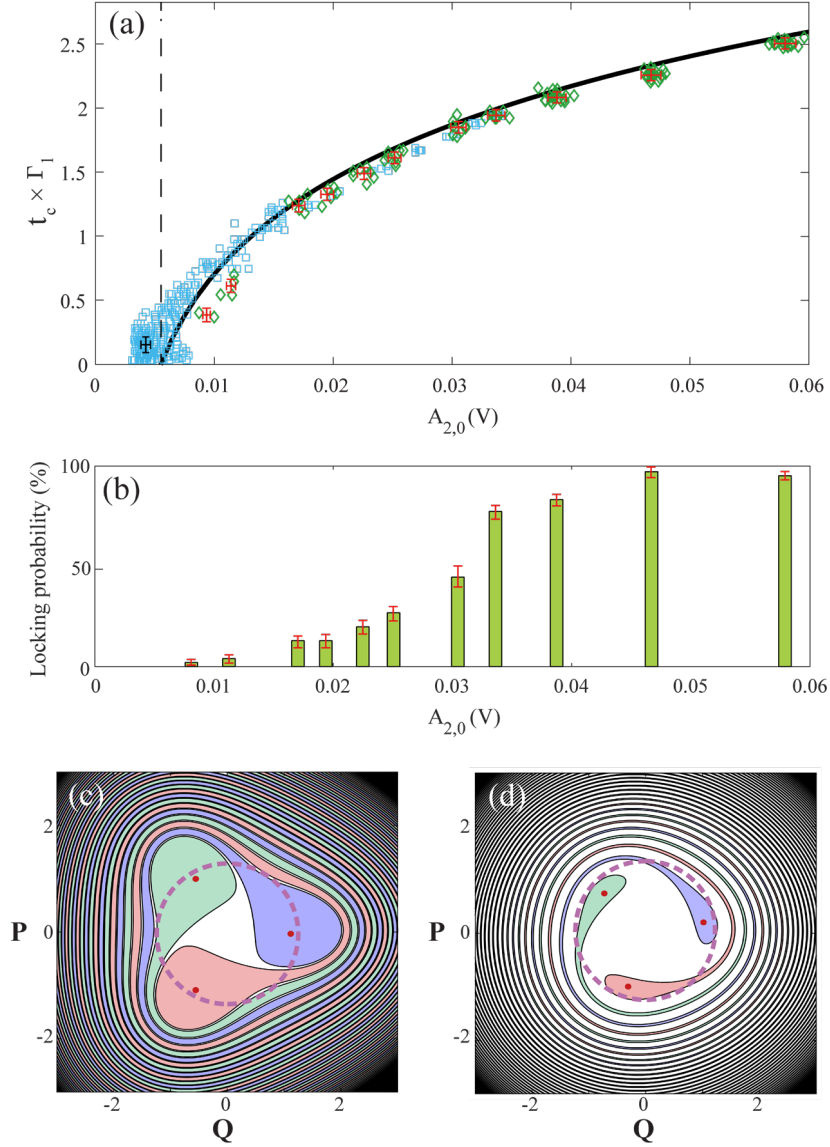

FIG. 4 Coherence time and the probability of locking. The coherence time and the locking probability increase as a function of the Mode 2 amplitude $A_{2,0}$ at the initial time of locking and are independent of all other parameters. (a) Measured $t_c$ in the units of Mode 1 decay time $1/\Gamma_1$. Blue squares are experiments preparing Mode 1 and Mode 2 at the phase-locked state initially, while green diamonds are measured when the modes are initially unlocked. The black dashed line $A_{2,threshold} \approx 0.006$ V is obtained from the averaged $A_{2,0}$ at $t_c \approx 0$ s. The black calculated coherence time is obtained via calculated the relaxation time of Mode 2 from $A_{2,0}$ to $A_{2,threshold}$ without using any adjustable parameters. The one standard deviation (SD) uncertainties for the blue squares are constant for each data point, labeled by the black cross, while the one SD uncertainties for each group of the green diamonds are labeled on the corresponding data groups [see Supplementary Note 6]. (b) The probability for locking increases as a function of $A_{2,0}$. The results are from 100 measurements with random initial relative phase $\phi_1 - \phi_2/3$ for each $A_{2,0}$. The uncertainties are one standard deviation of the corresponding Binomial distribution. (c) PTS basins of attraction for strong driving strength. (d) PTS basins of attraction for weak driving strength. The purple dashed rings represent the initial condition of Mode 1 with a fixed amplitude $A_{1,0}$ slightly above the PTS (red dots) and a random

relative phase. Starting points in the shaded (white) areas fall into the locked PTS states (unlocked trivial state). Larger Mode 2 amplitude increases the combined areas of the PTS basins of attraction. Compared to the experimental conditions, the schematic of the PTS separatrix uses a much smaller detuning term [see Supplementary Notes 3] for clarity of presentation. The experimental phase diagram is presented within Supplementary note 9, Figure S8.

## IV.   CONCLUSION AND OUTLOOK

A wide class of low-loss systems can be described by a set of independent nonlinear eigenmodes exchanging energy when two or more of their frequencies match, forming integer multiple relationships. In contrast to trivial brief energy exchanges occurring when the dynamically evolving eigenfrequencies intermittently match, here, using a simple, fully observable, two-mode system, we have shown how eigenmodes can phase-lock at internal resonances and remain locked for time periods up to 6x longer than the system relaxation times (i.e., $1/\Gamma_2$). This allows the modes to rapidly exchange large amounts of energy, producing non-monotonic modal energy evolution, making the locked system energy dissipation nonlinear, and resulting in completely different energy states compared to when the phase-locking has been bypassed. While locked, the energy exchange rate depends on the individual mode's eigenfrequency-energy relations and dissipation rates but does not significantly depend on the exact value of the coupling between them.

We have shown that entering or bypassing the phase-locked state depends sensitively on the initial relative phase, and thus the system's final state is a high gain nonlinear amplifier and discriminator of the relative phase and relative frequency. This phase dependence distinguishes the observed behavior from nonlinear damping recently observed in driven two-mode systems [27].

The experimentally measured relaxation dynamics of the two-mode system, including phase locking, is completely described by extending the well-understood model of a single mode system subject to a commensurate nonlinear drive, e.g., the period tripling state of a Duffing oscillator. The PTS model is accurate when the higher frequency mode dynamics are slow compared to the Mode 1 precession frequency in the PTS pseudopotential, i.e. when Mode 2 undergoes gradual energy change and the corresponding slow frequency shift according to its Duffing nonlinearity. This is clearly valid for our experiment and would be valid for many other systems with low natural decay rates. Additionally, for a system with small time-averaged interaction energy relative to the modes' energies, the long timescale energy evolution can be modeled by a single-parameter energy transfer equation (Eq. S6), providing an intuitive picture simplifying the general fully-coupled equation of motion description (Supplementary Note 8). While phase-locking is predicted within the PTS model, more general theoretical approaches [13] are needed to study the complete parameter range where the discovered phase-locking phenomenon may be observed outside the validity range of the PTS model.

Numerical simulation of our PTS model can qualitatively reproduce ~~and intuitively explain~~ the contradictory experiment results of the rapid energy loss [12], gain [11,21], and non-monotonic energy (this work) ~~for coupled states~~ using different effective Duffing coefficients of Mode 2 (Supplemental Note 4). However, we cannot confirm that our model is applicable to the specific experimental systems other than the one studied here. More complex phenomena could also lie at the root of such contradictory results. Within this model the direction of the energy exchange depends on the specifics of each mode's eigenfrequency-energy relationship near the internal resonance and can be the same or opposite to the one observed in the present experiments. For example, Mode 2 with a linear (zero) eigenfrequency-energy relationship can lock Mode 1 at a constant frequency and amplitude as shown in Fig 2, while Mode 2 with a hardening eigenfrequency-energy relationship can drag the locked Mode 1 to decay even faster than its natural relaxation rate, opposite to the energy gain effect shown in this work. Beyond the coupled Duffing oscillators, this model can be extended to describe two modes with arbitrary nonlinearities near an internal resonance (Supplementary Note 7).

During the preparation of this manuscript, the authors learned that the energy transfer between coupled modes at 1:3 internal resonance is studied by Y. Yan *et. al.* Complementing the present study of the transient relaxation, Yan *et.al.* independently demonstrate that the same period-tripling approach quantitatively describes the steady-state dynamics of such systems subject to a continuous external drive of the higher frequency mode. Demonstrating the rich dynamics possible in such systems, when the nonlinear second mode is driven sufficiently strongly, it responds with one of two distinct amplitudes. In combination with the period tripling for the lower mode, this leads to a total of 6 distinct discrete time-translation symmetry breaking states. Parametric 2*f* drive for mode 2 also results in 6 coexisting states.

A single nonlinear mode can be driven parametrically at arbitrary frequency multiples. Similarly, pairwise mode locking can occur at resonances of different integer fractions, other than 1:3 studies here. Moreover, the proposed description framework may be fruitfully applied not only to driven systems [Yan *et.al.*], but also to systems with more than pairwise resonances, i.e., when more than two modes (and/or drives) participate in the resonant energy exchange (Supplementary Note 7). While, generally, much richer, such systems may exhibit extended locked states, strongly influencing their dynamics under some conditions. Our work opens up a new window for exploration and harnessing of such a general class of nonlinear systems, both in and out of thermal equilibrium. Applying them to many-body systems with large numbers of coupled elements could lead to additional interesting applications, such as tailoring the energy flow direction and efficiency [9] and the broadband sensing with large assemblies of coupled oscillators [28]. These results are not limited to nonlinear mechanics, and shed light on coupled nonlinear resonator systems across physical domains, including, for example, nonlinearly coupled optical modes in chip-scale photonic resonators.

## ACKNOWLEDGMENTS
We thank Dr. Steven Shaw, Dr. Ori Shoshani and Dr. David A. Czaplewski for useful discussions. We thank Dr. Jason J. Gorman, Dr. Damián H. Zanette, Dr. J. Alexander Liddle, and Dr. Marcelo Davanco for reviewing this paper and giving meaningful suggestions. M.W. and D.J.P. are supported by the Cooperative Research Agreement between the University of Maryland and the National Institute of Standards and Technology Center for Nanoscale Science and Technology, Award 70NANB14H209, through the University of Maryland.

## APPENDIX A: PREPARATION OF THE SYSTEM'S INITIAL STATE

We use two methods for setting the initial condition of Mode 1 and Mode 2 in the ringdown experiment. The first method is the same as the one we used previously in the internal resonance research where we drive Mode 1 solely to the nonlinear regime and match the drive frequency $\omega_1$ to $\omega_2/3$ [11,21]. Due to resonantly enhanced energy transfer, energy flows between the two modes periodically, generating a limit cycle [29] [Supplementary Note 2]. In such a case, the two modes are prepared in the phase-locked state before ringdown, i.e., Mode 1 is at the PTS initially. By fine-tuning the turning-off time on the limit cycle, we can control the initial amplitude (energy) of Mode 2.

When the drive to Mode 1 at internal resonance is strong, the strong interaction between Mode 1 and Mode 2 terminates the limit cycle and both modes drop to the ground states. It limits the maximum $A_{2,0}$ that can be obtained in this way, and, therefore, the longer $t_c$, shown as blue squares in Fig. 4(a). To obtain even longer coherence time and explore larger parameter space, we use the second method that drives Mode 1 and Mode 2 via sequential sweeps of two simultaneously applied independent drives. We first sweep the driving frequency of Mode 1 to a value much higher than $\omega_2/3$ while keeping the drive of Mode 2 off. After reaching the high Mode 1 frequency, we maintain the Mode 1 drive while turning on the drive at Mode 2 frequency and sweep it down until Mode 2 reaches the desired initial amplitude. As Mode 2 has a lower frequency at high amplitude due to its spring softening effect, the frequency mismatch between the modes is large and interaction is weaker from the internal resonance. We then tune the driving frequency of Mode 1 down to the desired value. Two modes maintain their initial amplitudes until we start the free decay by simultaneously turning off the drives [black dots in Fig. 3(a)]. The initially uncoupled modes allow us to study the locking probability and directly observe the two different energy relaxation pathways of the system.

## APPENDIX B: PHASE DIAGRAM OF PERIOD-TRIPLING STATES

In this specific experiment, the mode nonlinearity can be well described by the Duffing nonlinearity and therefore, Eq. (1) is rewritten as:

$$\ddot{q}_1 + \Gamma_1 \dot{q}_1 + \omega_{1,linear}^2 q_1 + \alpha_1 q_1^3 = F_2 q_1^2 \cos(\omega_2 t). \tag{2}$$

It is convenient to write Eq. (2) in the rotating frame with a reference of $\omega_2/3$

$$q_1 = CQ \cos\left(\frac{\omega_2 t}{3}\right) + CP \cos\left(\frac{\omega_2 t}{3}\right) \tag{3}$$

where $P$ and $Q$ are two dimensionless quadratures and $C = \sqrt{\frac{8\omega_{1,linear}\delta\omega}{3\alpha_1}}$ with detuning $\delta\omega = \omega_2/3 - \omega_{1,linear}$. In the rotating frame, the dimensionless quasienergy can be written as [24]:

$$g(Q,P) = \frac{1}{4}(Q^2 + P^2 - 1)^2 - f(Q^3 - 3P^2 Q) \tag{4}$$

where $f = \frac{F_2}{3\sqrt{24\omega_{1,linear}\alpha_1\delta\omega}}$ is the dimensionless driving strength. The equation of motion for Mode 1 is written as:

$$\begin{aligned}\dot{Q} &= \partial_P g - \kappa Q \\ \dot{P} &= -\partial_Q g - \kappa P\end{aligned} \tag{5}$$

where $\kappa = \frac{\Gamma_1}{2\delta\omega}$. Trajectories in the phase diagram, as shown in Figs. 1(e), (f), are obtained by numerically integrating (5) from different initial conditions. The separatrices of the phase diagram are further calculated by doing gradient ascendant from the saddle points of $g(Q,P)$ based on Eq. (5) [26] [see Supplementary Note 3].

## REFERENCE

[1] F. Verstraete, M. M. Wolf, and J. Ignacio Cirac, *Quantum Computation and Quantum-State Engineering Driven by Dissipation*, Nature Physics **5**, 9 (2009).
[2] U. Weiss, *Quantum Dissipative Systems*, Vol. 13 (World Scientific, 2012).
[3] P. J. E. Peebles, *Physical Cosmology*, Vol. 67 (Princeton University Press, 2015).

[4] G. S. MacCabe, H. Ren, J. Luo, J. D. Cohen, H. Zhou, A. Sipahigil, M. Mirhosseini, and O. Painter, *Nano-Acoustic Resonator with Ultralong Phonon Lifetime*, Science **370**, 840 (2020).

[5] C. Audoin and B. Guinot, *The Measurement of Time: Time, Frequency and the Atomic Clock* (Cambridge University Press, 2001).

[6] V. B. Braginsky, V. P. Mitrofanov, and V. I. Panov, *Systems with Small Dissipation* (University of Chicago Press, 1985).

[7] I. Prigogine, *Non-Equilibrium Statistical Mechanics* (Courier Dover Publications, 2017).

[8] M. Onorato, L. Vozella, D. Proment, and Y. V. Lvov, *Route to Thermalization in the α-Fermi–Pasta–Ulam System*, Proceedings of the National Academy of Sciences **112**, 4208 (2015).

[9] R. I. Sosa and D. H. Zanette, *Energy Exchange in Globally Coupled Mechanical Phase Oscillators*, Phys. Rev. E **102**, 012208 (2020).

[10] A. W. Barnard, M. Zhang, G. S. Wiederhecker, M. Lipson, and P. L. McEuen, *Real-Time Vibrations of a Carbon Nanotube*, Nature **566**, 89 (2019).

[11] C. Chen, D. H. Zanette, D. A. Czaplewski, S. Shaw, and D. López, *Direct Observation of Coherent Energy Transfer in Nonlinear Micromechanical Oscillators*, Nat Commun **8**, 15523 (2017).

[12] J. Güttinger, A. Noury, P. Weber, A. M. Eriksson, C. Lagoin, J. Moser, C. Eichler, A. Wallraff, A. Isacsson, and A. Bachtold, *Energy-Dependent Path of Dissipation in Nanomechanical Resonators*, Nature Nanotech **12**, 631 (2017).

[13] O. Shoshani, S. W. Shaw, and M. I. Dykman, *Anomalous Decay of Nanomechanical Modes Going Through Nonlinear Resonance*, Sci Rep **7**, 18091 (2017).

[14] N. E. Bousse, J. M. Miller, H.-K. Kwon, G. D. Vukasin, and T. W. Kenny, *Quality Factor Tuning of Micromechanical Resonators via Electrical Dissipation*, Applied Physics Letters **116**, 023506 (2020).

[15] A. H. Nayfeh and D. T. Mook, *Nonlinear Oscillations* (John Wiley & Sons, 2008).

[16] I. Kozinsky, H. C. Postma, I. Bargatin, and M. L. Roukes, *Tuning Nonlinearity, Dynamic Range, and Frequency of Nanomechanical Resonators*, Applied Physics Letters **88**, 253101 (2006).

[17] K. L. Ekinci and M. L. Roukes, *Nanoelectromechanical Systems*, Review of Scientific Instruments **76**, 061101 (2005).

[18] J. Rieger, T. Faust, M. J. Seitner, J. P. Kotthaus, and E. M. Weig, *Frequency and Q Factor Control of Nanomechanical Resonators*, Applied Physics Letters **101**, 103110 (2012).

[19] Y. Tsaturyan, A. Barg, E. S. Polzik, and A. Schliesser, *Ultracoherent Nanomechanical Resonators via Soft Clamping and Dissipation Dilution*, Nature Nanotechnology **12**, 776 (2017).

[20] F. Mangussi and D. H. Zanette, *Internal Resonance in a Vibrating Beam: A Zoo of Nonlinear Resonance Peaks*, PloS One **11**, e0162365 (2016).

[21] D. Antonio, D. H. Zanette, and D. López, *Frequency Stabilization in Nonlinear Micromechanical Oscillators*, Nat Commun **3**, 1 (2012).

[22] L. Manevitch and A. I. Manevich, *The Mechanics Of Nonlinear Systems With Internal Resonances* (World Scientific, 2005).

[23] N. Lörch, Y. Zhang, C. Bruder, and M. I. Dykman, *Quantum State Preparation for Coupled Period Tripling Oscillators*, Phys. Rev. Research **1**, 023023 (2019).

[24] J. Gosner, B. Kubala, and J. Ankerhold, *Relaxation Dynamics and Dissipative Phase*

*Transition in Quantum Oscillators with Period Tripling*, Phys. Rev. B **101**, 054501 (2020).

[25] Y. Zhang, J. Gosner, S. M. Girvin, J. Ankerhold, and M. I. Dykman, *Time-Translation-Symmetry Breaking in a Driven Oscillator: From the Quantum Coherent to the Incoherent Regime*, Phys. Rev. A **96**, 052124 (2017).

[26] Y. Zhang and M. I. Dykman, *Nonlocal Random Walk over Floquet States of a Dissipative Nonlinear Oscillator*, Phys. Rev. E **100**, 052148 (2019).

[27] X. Dong, M. I. Dykman, and H. B. Chan, *Strong Negative Nonlinear Friction from Induced Two-Phonon Processes in Vibrational Systems*, Nature Communications **9**, 1 (2018).

[28] C. B. Wallin, R. De Alba, D. Westly, G. Holland, S. Grutzik, R. H. Rand, A. T. Zehnder, V. A. Aksyuk, S. Krylov, and B. R. Ilic, *Nondegenerate Parametric Resonance in Large Ensembles of Coupled Micromechanical Cantilevers with Varying Natural Frequencies*, Physical Review Letters **121**, (2018).

[29] S. Houri, D. Hatanaka, M. Asano, R. Ohta, and H. Yamaguchi, *Limit Cycles and Bifurcations in a Nonlinear MEMS Resonator with a 1:3 Internal Resonance*, Appl. Phys. Lett. **114**, 103103 (2019).

# Supplementary Information

**Persistent nonlinear phase-locking and non-monotonic energy dissipation in micromechanical resonators**

Mingkang Wang[1,2], Diego J. Perez-Morelo[1,2], Daniel Lopez[3,1], and Vladimir A. Aksyuk[1]

[1]Microsystems and Nanotechnology Division, National Institute of Standards and Technology, Gaithersburg, MD 20899 USA

[2]Institute for Research in Electronics and Applied Physics, University of Maryland, College Park, MD 20742, USA

[3]Materials Research Institute, Penn State University, University Park, PA 16802

## S1- Measurement Setup

In this work, we use a well-established doubly clamped beam micromechanical resonator as the experimental platform to study the energy decay processes of two nonlinearly coupled modes. The silicon resonator consists of a beam with lateral comb-drive electrodes for electrical actuation and detection [1,2]. In addition to the electrical measurement, we also perform optical measurement on the resonator out-of-plane motion as a separate measurement channel. Figure S1 shows a detailed schematic of the measurement setup. It is noteworthy to mention that the capacitive signal has a larger readout gain for the in-plane displacement since it is measured as the capacitive variance between the beam and the lateral electrode. On the contrary, the optical measurement, which is more sensitive to the out-of-plane displacement, was performed by a laser Doppler vibrometer (LDV) with a helium-neon laser ($\lambda$ = 633 nm) and maximum vertical detection limit of ±75 nm.

During the experiment, the doubly clamped beam is placed in a vacuum chamber of the pressure of $\approx$ 0.1 Pa ($\approx 10^{-3}$ Torr), at room temperature. Two a.c. drives of independently set frequency are applied on the side electrodes with a d.c. bias of 6 V. The capacitive (via a low-noise current amplifier) and the optical readout were simultaneously monitored with a lock-in amplifier and an oscilloscope. For the ringdown experiment, we switch off the drives and monitor the motion of the beam from the two readout channels of the oscilloscope whose bandwidth is set to be much larger than the relaxation rates of the modes.

The experimental setup developed in this study advances the conventional setups shown in previous works [3], giving us additional control and detection ability. On the one hand, it allows us to independently drive the modes of interest by combining electrostatic forces of different frequencies from the lateral electrodes in an open-loop configuration, making it possible to independently set the arbitrary desired initial condition of the modes. On the other hand, the two channels of detection enable us to detect the displacement of in-plane and out-of-plane modes capacitively and optically, respectively, for achieving the best readout gain and measurement sensitivity. With the appropriate calibration, it allows us to accurately measure the total energy of the coupled-modes-system which is not achievable before [2,3].

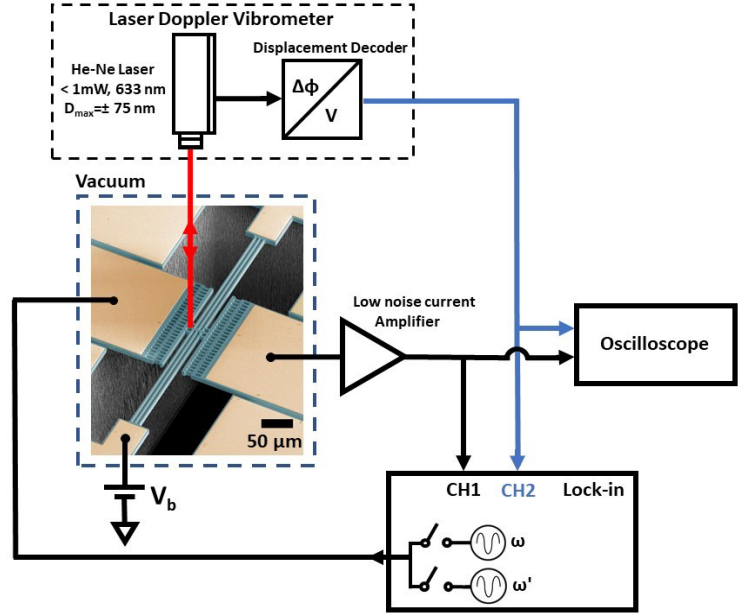

**Figure S1. Measurement setup**: The schematic shows the clamped-clamped beam resonator and the wiring of the capacitive (black) and optical (blue) detection. In addition to the electrical measurement through the lateral comb electrode, we measure the out-plane mode using a laser Doppler vibrometer.

## S2 Two modes of interest and preparation of their initial states

We consider two mechanical modes on the doubly clamped beam. They initially have nearly 1:3 commensurate eigenfrequencies. Figures S2 (a) and (b) show the measured linear resonance of Mode 1 and Mode 2. As discussed in the previous section, we measure the displacement of the modes both electrically and optically, shown as the blue and orange dots. The Lorentzian fit of the mode shows $\omega_{1,linear}/2\pi \approx 64630.6$ Hz, $\Gamma_1/2\pi \approx 1.5$ Hz, and $\omega_{2,linear}/2\pi \approx 199881.8$ Hz, $\Gamma_2/2\pi \approx 3.3$ Hz. With increasing the driving force, Mode 1 turns into a nonlinear regime shown as Fig. S2(c) where a spring hardening effect is observed (positive Duffing coefficient). When the a.c. voltage is larger than 0.06 V, the Duffing curve of Mode 1 reaches the internal resonance ($\omega_1:\omega_2 = 1:3$) during the upward frequency sweep. Due to the strong nonlinear intermodal coupling, it cannot pass the internal resonance with Mode 2 unless the driving voltage is higher than 0.45 V [1,2]. For such voltage Mode 1's Duffing curve can be swept to higher frequencies shown as the red line for example. Unlike Mode 1, Mode 2 shows the spring softening effect (negative Duffing coefficient) as shown in Fig. S2(d). The fit based on Duffing nonlinearity (black line) describes well the electrically/optically measured mechanical response spectrum of Mode 2.

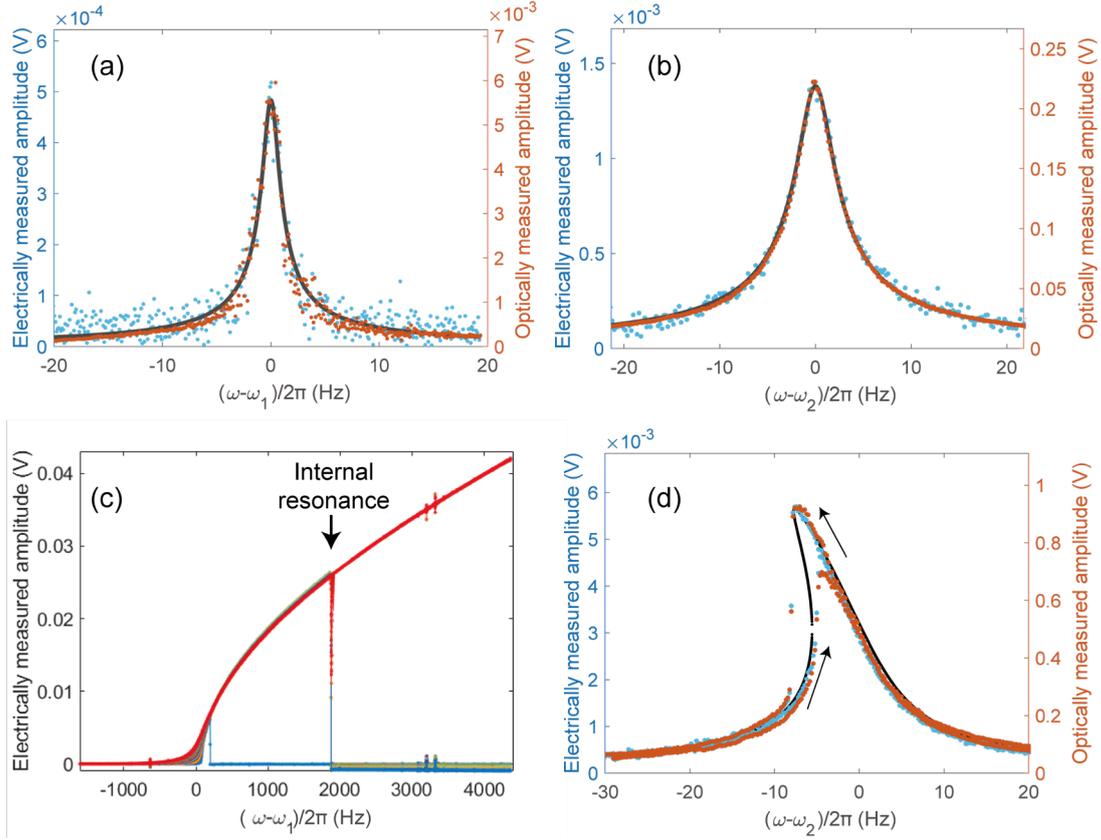

**Figure S2**. **Two mechanical modes with nearly 1:3 commensurate frequency relationship**: (a) Linear, small-amplitude resonance of Mode 1. The blue and orange dots present the measured data from the electrical and optical detection, respectively. The black line is from the Lorentzian fit. (b) Linear, small amplitude resonance of Mode 2. (c) Nonlinear response of Mode 1 for the a.c. component of the diving voltage ranging from 0.01 V to 0.81 V with a 0.05 V increment. (d) Nonlinear resonance of Mode 2 with a hysteresis. The black line is a Duffing model fit. The arrows label the sweeping directions.

To monitor the energy decay process of the two modes, we need to prepare the two modes to preset initial conditions. We used two methods to perform the preparation of the initial state. In the first set of experiments [Fig. 2 in the main text], we prepare the two modes at internal resonance [black arrow in Fig. S2 (c)] by only driving Mode 1. The drive frequency is swept up to the internal resonance at $\omega_2/3$. The strong nonlinear intermodal coupling transfers energy between Mode 1 to Mode 2 back and forth, inducing limit cycles. Figures S3 (a) and (b) present the time-varying amplitude of Mode 1 and Mode 2, respectively, at a constant drive amplitude and a fixed frequency near the internal resonance. The amplitudes of Mode 1 and Mode 2 periodically vary, with short time intervals when energy is exchanged quickly (faster than ringdown times) back and forth between them. We turn the drive to Mode 1 off at a specific time, for example, as labeled by the black dashed line. The initial conditions for the two modes at the beginning of the free ringdown are the corresponding values of amplitude at that time. It is noteworthy that although we only drive Mode 1, the maximum energy store in Mode 2 can exceed Mode 1 energy by up to 10 times at certain points in the cycle shown in Fig. 2(a). In the second set of experiments [Fig. 3 in the main text], we separately drive Mode 1 and Mode 2 by two independently controlled drives, by which we can set the initial conditions of the two modes to whatever condition we want. However, later experimental results show that the coherence time only strongly depends on the amplitude of Mode 2 at the beginning of locking ($A_{2,0}$).

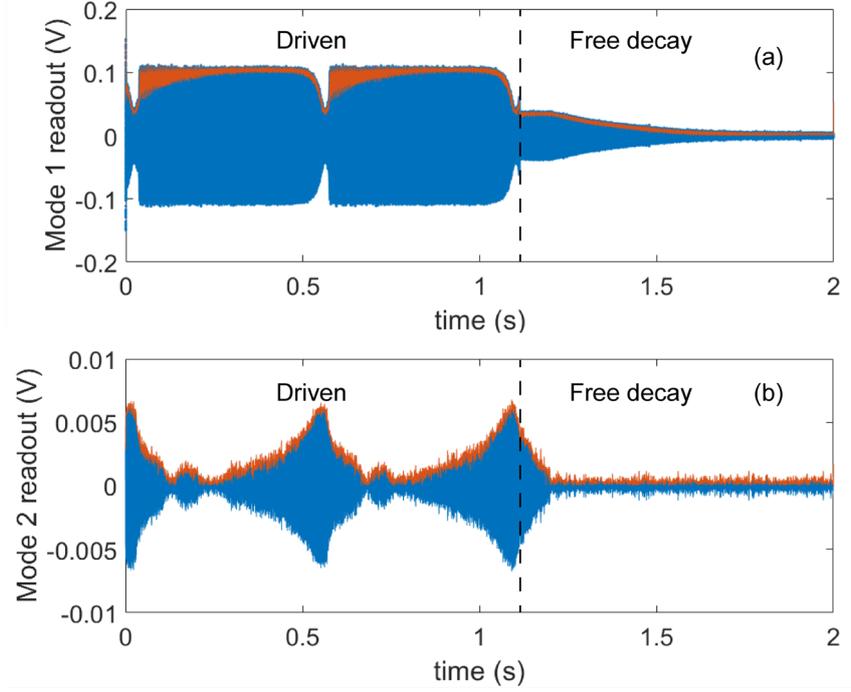

**Figure S3**. **Preparation of the initial state at the internal resonance.** Electrically measured displacement of (a) Mode 1 and (b) Mode 2. A limit cycle under a steady drive is followed by the free decay of the coupled system after the drive is turned off (black dashed line) near Mode 2's maximum amplitude. The orange line labels the vibration amplitude. The electrical capacitive measurement gain for torsional Mode 2 is much smaller than for translational Mode 1.

## S3- Simulation of Period tripling states

As shown in Eq. (1), Mode 1 performs as an anharmonic oscillator with resonance frequency dependent on amplitude, and it is subject to the external excitation from Mode 2 via the nonlinear coupling term $\propto q_1^2 q_2$. In this specific experiment, the amplitude-frequency relationship for Mode 1 is well described by the Duffing form, therefore, Eq. (1) is rewritten as:

$$\ddot{q}_1 + \Gamma_1 \dot{q}_1 + \omega_{1,linear}^2 q_1 + \alpha_1 q_1^3 = F_2 q_1^2 \cos(\omega_2 t) \qquad (S1)$$

where $\alpha_1$ stands for the Duffing coefficient and $\omega_2 \approx 3\omega_1$ where $\omega_1^2(q_1) = \omega_{1,linear}^2 + \alpha_1 q_1^2$ with detuning $\delta\omega = \omega_2/3 - \omega_{1,linear}$. The product of the $\cos(\omega_2 t)$ term and the approximately-harmonic displacement term $q_1^2 \propto \cos(\omega_2 t/3)^2$ provides a driving force with a harmonic component at the frequency $\omega_2/3 = \omega_{1,linear} + \delta\omega$ to drive Mode 1.

Following Ref [4], it is convenient to write Eq. (S1) in the rotating frame with a reference frequency of $\omega_2/3$

$$q_1 = CQ \cos\left(\frac{\omega_2 t}{3}\right) + CP \cos\left(\frac{\omega_2 t}{3}\right) \qquad (S2)$$

where $P$ and $Q$ are two dimensionless quadratures and $C = \sqrt{\frac{8\omega_{1,linear}\delta\omega}{3\alpha_1}}$. In the rotating frame, the

dimensionless quasienergy can be written as:

$$g(Q,P) = \frac{1}{4}(Q^2 + P^2 - 1)^2 - f(Q^3 - 3P^2Q) \tag{S3}$$

where $f = \frac{F_2}{3\sqrt{24\omega_{1,linear}\alpha_1\delta\omega}}$ is the dimensionless driving strength. The equation of motion for Mode 1 is written as:

$$\begin{aligned}\dot{Q} &= \partial_P g - \kappa Q \\ \dot{P} &= -\partial_Q g - \kappa P\end{aligned} \tag{S4}$$

where $\kappa = \frac{\Gamma_1}{2\delta\omega}$. Trajectories in the phase diagram [as shown in Fig. 1e] can be obtained by numerically integrating (S4) from different initial conditions.

By solving for $\dot{Q} = 0$ and $\dot{P} = 0$, we obtain 7 solutions. Then we use $\ddot{Q} < 0$ and $\ddot{P} < 0$ to select the 3 stable states (red dots in separatrix plot), 3 saddle points, and 1 trivial-state.

The separatrix in the P-Q phase diagram is on the ridge of $g(P,Q)$ which separates the phase diagram into 4 regimes. To obtain the separatrix, we set the initial condition P(0), Q(0) at around the saddle points which are on the minimum point of a ridge of $g(P,Q)$, and do gradient ascendant of P, Q, i.e. inverting time in (S4) and numerically integrating, as:

$$\begin{aligned}Q(t+dt) &= Q(t) - (\partial_P g - \kappa Q)dt \\ P(t+dt) &= P(t) - (-\partial_Q g - \kappa P)dt.\end{aligned} \tag{S5}$$

The trajectory of P(t)-Q(t) is the separatrix.

Note, the present separatrixes in the paper have $\kappa = 0.1$, while our real system losses are much smaller, $\kappa < 0.001$. We use bigger $\kappa$ for visual clarity of the diagrams presented in our paper, to avoid the helix of separatrix winding too dense and narrow. The performance of the system is qualitatively the same for different $\kappa$. In order to show the separatrix under different driving strength, we set medium $f = 0.0800$ for Fig. 1(e) and the red line in Fig. 2(f), and small $f = 0.0034$ for Fig. 1(f) and the blue line in Fig. 2(f). Large $f = 0.25$ and medium $f = 0.05$ are used for Fig. 4 (c) and (d), respectively.

To illustrate the evolution of an oscillator under a slowly time-varying driving strength (mimicking the gradual amplitude decay of Mode 2), we generate a movie (Supplementary Movie 1) where the oscillator is starting from two different initial conditions having the same amplitude but slightly different phases (<0.1 rad). For the Movie, $\kappa = 0.1$ is set as a constant while $f$ linearly changes from 0.25 to 0 during the ringdown. In the beginning, the oscillator freely decays. The red dot depicts the trajectory of the oscillator with an initial phase that allows it to lock into the PTS, while the blue one shows the trajectory that bypasses the PTS and decays to the trivial state directly. At the time when the gradually decreasing $f$ reaches a threshold $f_0 \approx 0.033$, the PTS attractors disappear. The oscillator (red dot) unlocks from the PTS and rings down to the trivial state. The simple PTS model qualitatively presents the dynamics of Mode 1 observed in Figure 2.

The same dynamics are also depicted in Figure S4, presenting the amplitude as a function of time during the decay from the two initial conditions, numerically integrating (S4) with $f$ gradually decreased in time, as noted above. The sharp amplitude drop visible when the resonance is being bypassed (blue) is consistent with the measurement in Fig. 3c. The linear-damping relaxation rate is recovered once the modes are unlocked. Note, here the frequency $\omega_2$ is maintained constant while $f$ is decreasing, therefore, the locked state (red) shows a constant amplitude. This is consistent with the measurements in Fig. 2a, where Mode 2's frequency is nearly unchanged (Fig. 2d). The dynamics of the PTS under time-varying modulation frequency and force is more fully considered in the next Supplemental section, showing Mode 1 exhibiting a non-monotonic energy change.

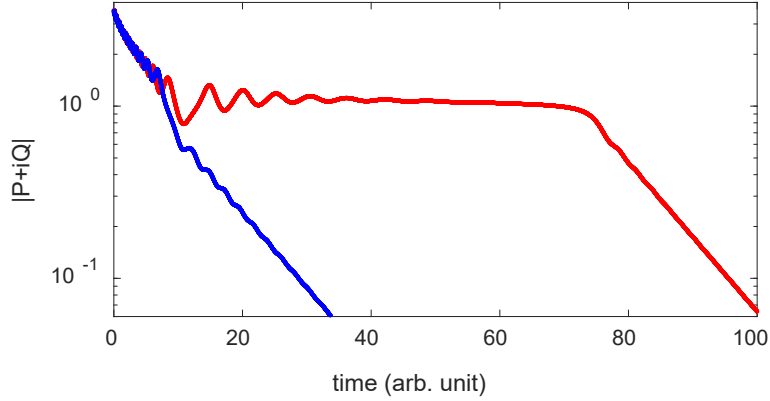

**Figure S4 Amplitude of the oscillators shown in Supplementary Movie 1.** The red and blue lines correspond to the dots in Supplementary Movie 1 with the corresponding colors.

## S4- PTS under dynamically varying parameters

As we discussed in the previous section, Mode 1 can be regarded as an oscillator subject to PTS, governed by the Eq. (S1), where Mode 2 provides the external drive with frequency $\omega_2 \approx 3\omega_1$ and force $F_2 \propto q_2$, whenever any fast dynamics of Mode 2 due to the coupling can be neglected. This simple model not only can qualitatively explain the behavior of Mode 1 in this experiment, such as the locking and bypass trajectories [illustrated in Fig.1 (e), (f); Fig. S4], the locking probability [Fig 4(c), (d)] and its striking energy decay reversal. The model can also provide a common general approach for describing a multitude of different experimentally observed energy-dependent dissipation rates in systems at internal resonance [1-3], accounting for both an increased as well as decreased and even negative dissipation in the locked state.

Figure S4 uses the PTS model to explain how PTS locks an oscillator [Mode 1] and makes it persist without decay for a period of time (coherence time), longer than its intrinsic dissipation time, during which the driving force is gradually decreasing. It describes the behaviors of Mode 1 in Fig. 2(a) where Mode 2 does not show strong nonlinearity, i.e., the driving frequency is nearly constant $\omega_2 \approx \omega_{2,linear}$. In order to explain the non-monotonic dissipation rate showing in Fig. 3(c) by this PTS model, we need to let the driving frequency from Mode 2 $\omega_2$ vary together with the force $F_2$, so that we can take into account the amplitude-dependent frequency of nonlinear Mode 2. As the reference frequency $\omega_2/3$ of the rotating frame is also changing in time in this case, we use the stationary frame to simulate the dynamics [i.e. directly numerically integrating Eq. S1] and extract the amplitude of the oscillation to avoid any additional approximations or ambiguity. However, we note that this is equivalent to slowly changing variables $F_2$ and $\omega_2/3$ in the rotating frame, which now rotates at a variable rate.

The equation of motion is shown as Eq. (S1) where $\omega_{1,linear}/2\pi = 10$ kHz, $\Gamma_1/2\pi = 0.1$ Hz (quality factor $= \omega_{1,linear}/\Gamma_1 = 1.0 \times 10^5$ ), $\alpha_1 = 1.0 \times 10^4$ rad$^2$ s$^{-2}$ nm$^{-2}$ are used for this example. In our paper, Mode 2 has a negative Duffing nonlinearity, therefore, $\omega_2$ increases following the gradual decrease of $q_2 \propto F_2$ to mimic the energy loss of nonlinear Mode 2. The blue lines in Figure S5(b),(c) shows the time-dependent $F_2$ and $\delta\omega$ that we used in the simulation where

$F_2$ decreases from $1.0 \times 10^5$ rad$^2$ s$^{-2}$ nm$^{-1}$ to $0$ rad$^2$ s$^{-2}$ nm$^{-1}$ between $t = 0.1$ s and $t = 0.33$ s and $\delta\omega/2\pi$ increases from 100 Hz to 150 Hz between $t = 0.1$ s and $t = 0.2$ s. In Fig. S5(a), the oscillator is locked to the stable PTS between $t = 0$ s and 0.1 s. As $\omega_2$ increases, the oscillator exhibits an energy gain, qualitatively similar to the one observed in our experiment shown in Fig. 3(c). With decreasing $F_2$, it unlocks from PTS at $t \approx 0.2$ s where $F_2 < 0.6 \times 10^5$ rad$^2$ s$^{-2}$ nm$^{-1}$ and exponentially decays with its intrinsic dissipation rate afterward. The non-monotonic amplitude (energy) clearly indicates that the drive pumps energy into the oscillator. Intuitively, if we treat the period-3 drive as a perturbation, the oscillator (Mode 1) with a positive $\alpha_1$ follows its hardening eigenfrequency-energy dependence, where the amplitude increases with the locked increasing frequency. The sweeping-up frequency of the drive (Mode 2) drags the phase-locked oscillator (Mode 1) to a higher frequency, leading to a larger amplitude (pumping energy into the oscillator, gain effect). This simple simulation qualitatively explains the novel non-monotonic dissipation rate observed in our experiment.

More generally, the PTS model can also explain the rapid energy loss effect for oscillators at internal resonance, as has been recently observed experimentally in another system [3]. In this case, the nonlinear drive pumps energy out from the oscillator in the locked state. For this numerical experiment, the parameters of the oscillator remain the same, we only reverse the sign of $\delta\omega$, making $\omega_2$ decrease with decreasing $F_2$, shown as the red lines in Fig. S5(b),(c). As a result, the oscillator decays faster than its intrinsic dissipation rate while locked in the PTS with time-varying driving parameters [between approximately 0.1 s and 0.2 s]. The faster decay rate compared to its intrinsic decay rate clearly shows that the time-varying drive pumps energy out of the oscillator locked to the PTS in this case.

It is noteworthy that the rapid energy gain/loss is mainly due to the time-varying detuning which drags the locked mode. The time-varying nonlinear force does not contribute much to the amplitude change. The black line in Fig. S5(a) shows the case where the detuning $\delta\omega/2\pi = 100$ Hz is constant, while the driving force $F_2$ decreases following the same trend as before (black lines in Fig. S5(b),(c)). Here the amplitude of the mode does not show rapid changes. It shows a similar trend as Fig. S4 where the mode maintains a nearly constant amplitude during locking ($t \approx 0.1 - 0.2$ s) even though the force is decreasing. After unlocking at $t \approx 0.2$ s where $F_2 < 0.6 \times 10^5$ rad$^2$ s$^{-2}$ nm$^{-1}$, the mode begins to lose energy with its intrinsic loss rate.

Although the different behaviors at internal resonance shown in Ref. [2,3] (main text Ref. [11, 12]) and present work are also observed within our PTS model by simply changing the sign of Mode 2's Duffing coefficient, we do not have enough information on whether this explanation applies to Ref. [3].

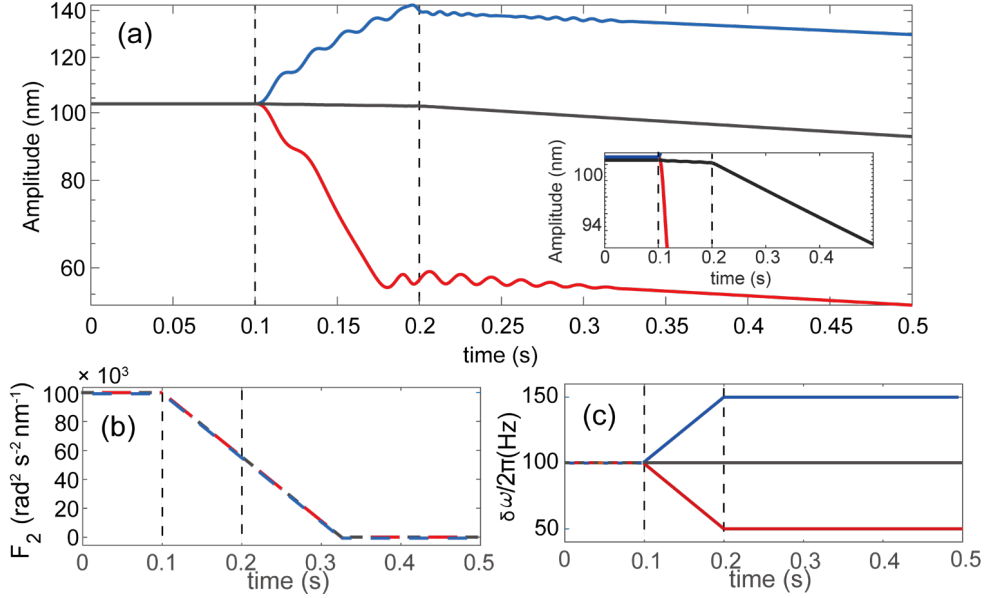

**Figure S5 Numerically modeled dynamics of oscillators in PTS with time-varying parameters.** $\omega_{1,linear}/2\pi = 10$ kHz, $\Gamma_1/2\pi = 0.1$ Hz, $\alpha_1 = 1.0 \times 10^4$ nm$^2$/s$^2$ are fixed. $F_2$, $\delta\omega$ change as shown in (b), (c), resulting in the ringdown shown in (a) marked with the corresponding colors. The opposite tuning direction of driving frequency between $t = 0.1$ s and $t = 0.2$ s drags the locked mode to a higher/lower oscillation frequency, exhibiting rapid energy gain/loss faster than its intrinsic dissipation rate.

## S5- Energy transfer between modes and calibration

Since the time-averaged interaction energy is neglected, the conservation of energy during the coherence time equals the energy lost by Mode 2 to the energy gained by Mode 1 after accounting for their intrinsic losses. It allows us to simultaneously fit both mode amplitudes, calibrate them, and plot them on a common energy scale.

After turning off the external driving force, the amplitude of Mode 1 remains constant or even increases for a finite period of time, and the energy of Mode 2 decays much faster than its expected exponential relaxation via linear damping. The equations describing the amplitude of each mode during ringdown for a coupled two-modes system at internal resonance are:

$$\frac{dE_1}{dt} = -\Gamma_1 E_1 + W$$

$$\frac{dE_2}{dt} = -\Gamma_2 E_2 - W \qquad (S6)$$

where $E_1 = c_1 A_1^2$ and $E_2 = c_2 A_2^2$ are the energies of Mode 1 and Mode 2, which are proportional to their measured amplitude square with proportionality constants of $c_1$ and $c_2$, respectively, and $W$ is the energy exchange rate between the two modes, which is approximately constant in our system, since $\frac{dE_1}{dt} \approx 0$ during the coherence time.

By fitting the ringdown amplitude of the two modes with Eq. (S6), we obtain the calibration constant $\eta = \frac{W/c_1}{W/c_2} = \frac{c_2}{c_1}$ which is used for unifying the unit of energy of Mode 1 and Mode 2.

By solving Eq. (S6), we obtain the time-dependent amplitude of Mode 1 and Mode 2:

$$A_1^2 = A_{1,0}^2 e^{-\Gamma_1 t} + \frac{W}{c_1 \Gamma_1}(1 - e^{-\Gamma_1 t})$$

$$A_2^2 = A_{2,0}^2 e^{-\Gamma_2 t} - \frac{W}{c_2 \Gamma_2}(1 - e^{-\Gamma_2 t}) \quad (S7)$$

where $A_{1,0}$ and $A_{2,0}$ are the initial amplitude of Mode 1 and Mode 2 at the beginning of locking, respectively. These equations were used to fit the data of Figure 2(a) and 3(e) with $W/c_2$ as the only fitting parameter (black lines in the figure). Without loss of generality, the energy units are chosen such that $W/c_1 = 1$. After unlocking, the two modes dissipate energy at the rate given by their intrinsic dissipation rate $\Gamma_1$ and $\Gamma_2$, as expected. The blue and red dashed lines in Fig. 2(a) indicate the independent exponential decays of the modes ($E_1 = E_{1,0} e^{-\Gamma_1 t}$, $E_2 = E_{2,0} e^{-\Gamma_2 t}$).

As Mode 2 is regarded as the driving force to Mode 1 to support the PTS, we can solve Eq. (S7) to get a coherence time $t_c$ by considering the ringdown time of mode 2 from its initial amplitude to the threshold of $A_{2,threshold}(f_0) \approx 0.006\ V$ as measured in Figure 4(a). With known $\Gamma_2$, $W/c_2$, and $A_{2,0}$, the coherence time can be written as:

$$t_c = -\frac{1}{\Gamma_2} \ln\left[\frac{A_{2,threshold}^2 + \Gamma_2 W/c_2}{A_{2,0}^2 + \Gamma_2 W/c_2}\right]. \quad (S8)$$

## S6- Uncertainty analysis

There are two groups of data (blue squares and green diamonds) shown in Figure 4(a), distinguished by how their initial condition are prepared. Their error bars are obtained separately for the two data types via slightly different methods.

For the case where Mode 1 is locked to Mode 2 initially (blue squares), the coherence time $t_c$ is defined as shown in Figure 2(e), where the start of coherence time is well-defined as the time we turn off the drive ($t = 0$). The end of coherence time is defined as the point of unlocking, and is determined with a larger uncertainty due to the oscillation of frequency. This is the leading uncertainty in $t_c$ for these data. For quantifying it, we use one period of oscillation at unlocking, shown as positive and negative error bars (y-axis) for each data point. The error bar of $A_{2,0}$ (x-axis) indicates one standard deviation due to the detection noise of the $A_2$ measurement at the beginning of the ringdown (when the drive is turned off). The measurement noise is constant and the differences in the oscillation frequency are small, thus the data points share approximately the same horizontal and vertical uncertainties, as indicated by the black error bars in Fig. 4(a).

For the case where Mode 1 is not locked to Mode 2 initially (green diamonds), the coherence time is defined as indicated in Figure 3(b), where both the start and the end of locking are uncertain due to the frequency oscillation. The uncertainty of $t_c$ is defined as an average of the oscillation periods at the start and the end of locking. Regarding the uncertainty of $A_{2,0}$, it is a combination of two statistically independent contributions. The first part is from the detection noise of $A_2$, the same as above. The second part is arising from the uncertainty at the starting point of the coherence period (locking time) due to the oscillation of frequency, as shown in Fig. 2(b). We define the peak-to-valley difference of $A_2$ during the first frequency oscillation period after locking as the second part of the uncertainty. We combine these two parts of uncertainty for $A_{2,0}$ via room-mean-square and depict as the error bars in the *x*-axis for the green dots. The oscillation of frequency is initially relatively large and amplitude and frequency vary between data, we characterize the uncertainties separately for each measurement, shown as the red error bars on top of the corresponding groups of green diamond data points shown in Fig. 4(a).

## S7- Nonlinear coupled modes in resonance

We conjecture that the model we have presented may provide a useful approximate description for a broad range of systems near integer fraction resonances. Specifically, many such systems may exhibit stable locked states between two nonlinear coupled modes, whenever at least one of them is described by an amplitude-dependent eigenfrequency. The direction of the energy flow needed to maintain the locked state is determined by the modes' eigenfrequency-amplitude dependences (as illustrated in the numerical experiment in Section S4 and further discussed below). The stability of the locked state for different signs of these dependencies, as well as the conditions for applicability of the adiabatic assumption, are interesting subjects for future research, and may be studied further using existing general theoretical approaches [6], however, the following consideration may serve as a useful guide.

In our simplified model, we make a few assumptions. We start by assuming the energy of the system is predominantly described by the sum of the energies of the two modes, neglecting the time-averaged energy of the interaction term. We further neglect the higher harmonics and describe each mode by its harmonic eigenfrequency being dependent on its energy. Duffing oscillator is just one specific example of such an energy-eigenfrequency relationship. Closely related is the assumption that the modes are constrained to their eigenfrequency-amplitute relationships, largely independent of the motion of the other mode: neglecting the time-averaged interaction energy makes the eigenfrequency-energy relationships for each mode independent of the energy of the other mode. These frequency-energy relationships $\omega_{k,osc}(E_k); k = 1,2$ are the first cornerstone of the proposed model.

Within this description, modes are allowed to exchange energy when their frequencies are commensurate. Energy conservation, after accounting for independent dissipation by each mode, is the second cornerstone of the model,

$$\frac{dE_1}{dt} = -\Gamma_1 E_1 + W(t) \qquad (S9)$$

$$\frac{dE_2}{dt} = -\Gamma_2 E_2 - W(t)$$

where $W(t)$ is the energy exchange rate (power flow) between the modes, which is time-dependent, generally. (Cf. Eq. S6, where power flow is near-constant for our specific experimental system).

The third cornerstone is the persistent phase-locked state on resonance, similar to the period tripling state for a single nonlinearly-driven Duffing oscillator, with the higher frequency mode acting as the nonlinear drive. The final key simplification here is the assumption that the energy, and therefore also the frequency, of the higher frequency mode is changing sufficiently slowly, such that the stable states of period tripling states driven by the higher frequency mode evolves adiabatically compare to the dynamics of lower frequency mode. This specifically assumes the interaction with Mode 1 does not result in significant dynamical changes to Mode 2 frequency and amplitude other than due to the gradual energy gain/loss of Eq. (S9). This assumption is valid under our specific experimental conditions, with only small deviations, i.e., the slight *amplitude* oscillations observed in the Mode 1 and Mode 2 data in Figure 3c,e, and a small frequency 'pull' near the unlocking point if Figure 2d. Importantly, the Mode 2 *phase* remains constant or varies very smoothly everywhere except near the unlocking point. The observed Mode 1 phase oscillations are expected for an oscillator settling into its PTS following our model. The exact general conditions for the validity of this assumption remain to be investigated, such as by considering the perturbation from the small Mode 1 motion near the PTS on Mode 2, and the corresponding perturbation force back on Mode 1 using the more general coupled description [6].

The PTS-like locking to an external drive $F\cos(\omega_2 t)$ at frequency $\omega_2 \approx n\omega_1$ is not limited to a specific value of $n = 3$, but may exist for arbitrary integer $n \geq 1$, provided a nonlinear drive term is present, of the form $Fq_1^{n-1}\cos(\omega_2 t)$. Such effective drive term may arise from a nonlinear interaction or from a linear interaction with a nonlinear resonator mode (i.e., linear drive at the $n$-th harmonic). Here $n = 1$ is a simple linear drive on resonance, resulting in one non-zero state, and $n = 2$ is the common parametric drive at twice the frequency, resulting in two non-zero states, etc. As has been noted by others [5], the PTS at $n = 3$ is distinct from the lower $n$. For example, the stable trivial state co-exists with the PTS states over a frequency range.

The existence of such symmetry-breaking states for $n \geq 2$ is not limited specifically to Duffing nonlinearity, but rather is the property of any oscillator with an energy-dependent eigenfrequency $\omega_1(E_1)$ near the internal resonance, with the energy $E_1 = E_{1,res}$, where $E_{1,res}$ is defined by the period-n driving frequency $\omega_2$, i.e. $\omega_1(E_{1,res}) = \omega_2/n$. When the drive is applied by a higher-order coupled mode (Mode 2), we have $\omega_1(E_{1,res}) = \omega_2(E_{2,res})/n$. Since we are interested in describing systems near internal resonances, we can linearize the $\omega_k(E_k); k = 1,2$:

$$\omega_k(E_k) = \omega_k(E_{k,res}) + (E_k - E_{k,res})\frac{d\omega_k}{dE_k}(E_{k,res}). \qquad (S10)$$

Since the PTS ($n$ = 3) for a Duffing oscillator have been extensively studied, it is useful to reduce the motion of any nonlinear mode with $\frac{d\omega}{dE}(E_{res}) \neq 0$ sufficiently close to a PTS at $\omega_{res} = \omega(E_{res})$ to the equivalent Duffing oscillator having the same $E_{res}$, $\omega(E_{res})$ and $\frac{d\omega}{dE}(E_{res})$. Here we derive expression (S10) for the Duffing oscillator and explicitly provide the equivalent Duffing PTS parameters $\omega_0$, $\alpha$ and $\delta\omega$ for a mode with given $E_{res}$, $\omega(E_{res})$ and $\frac{d\omega}{dE}(E_{res})$.

For a Duffing oscillator with a linear resonance frequency of $\omega_0$ and a Duffing coefficient $\alpha$, the eigenfrequency-energy relationship is given by $\omega^2 = \omega_0^2 + \alpha\langle q^2\rangle$ and $E = K\langle q^2\rangle$, accounting for both kinetic and potential energies, with a stiffness $K = m\omega_0^2$ for a modal mass $m$. To simplify, we assume the $q$, energy, and force are scaled such that $m = 1$. Assuming $\alpha\langle q^2\rangle \ll \omega_0^2$ we have $\omega(E) = \sqrt{\omega_0^2 + \frac{\alpha}{\omega_0^2}E} \approx \omega_0 + \frac{\alpha}{2\omega_0^3}E$, or, in the form of Eq (S10),

$$\omega(E) \approx \omega_0 + \delta\omega + \left(E - 2\frac{\omega_0^3}{\alpha}\delta\omega\right)\frac{\alpha}{2\omega_0^3}. \qquad (S11)$$

It is evident that for a nonlinear mode near a PTS with given $E_{res}$, $\omega(E_{res})$ and $\frac{d\omega}{dE}(E_{res})$, the equivalent Duffing PTS parameters are $\delta\omega = E_{res}\frac{d\omega}{dE}(E_{res})$, $\omega_0 = \omega(E_{res}) - E_{res}\frac{d\omega}{dE}(E_{res})$, and $\frac{\alpha}{2\omega_0^3} = \frac{d\omega}{dE}(E_{res})$. The equivalent Duffing oscillator loss rate remains the same as the nonlinear mode loss.

Now, we consider again our model where an adiabatically-varying higher-frequency mode drives the lower-frequency mode of energy-dependent eigenfrequency. In a simple case of linear Mode 2, $\frac{d\omega_2}{dE_2} = 0$, we have the familiar PTS-like state with a constant-frequency nonlinear drive from Mode 2. Here we note that for Duffing Mode 1 changing a stiffening nonlinearity with a positive detuning to a softening nonlinearity with a negative detuning retains the same PTS but with a mirrored phase diagram in the $\omega_2/3$ rotating frame (Figs. 1e, f). More generally for modes of arbitrary energy-frequency relationship, the mirrored phase diagram will also present in the vicinity of the internal resonance upon changing $\frac{d\omega_1}{dE_1} \to -\frac{d\omega_1}{dE_1}$. In the rotating frame, near the resonance, the Mode 1 frequency-amplitude dependence is fully defined by

$$\omega_1(E_1) - \frac{\omega_2}{n} = (E_1 - E_{1,res}) \frac{d\omega_1}{dE_1} \tag{S12}$$

and the $\left|\frac{d\omega_{1,osc}}{dE_1}\right|$, together with the loss rate, describes Mode 1 dynamics near the internal resonance when dynamic backaction of Mode 1 on Mode 2 can be neglected.

Whenever the locked state $\omega_1(E_1) = \omega_2(E_2)/n$ is maintained in time, by taking a time derivative we obtain

$$\frac{dE_1}{dt} \frac{d\omega_1}{dE_1} = \frac{1}{n} \frac{dE_2}{dt} \frac{d\omega_2}{dE_2}. \tag{S13}$$

For known $\frac{d\omega_1}{dE_1}$ and $\frac{d\omega_2}{dE_2}$, Eq (S9) and (S13) can be solved to find the energy exchange rate $P(t)$, the time evolution of the system energies $E_k(t)$ and, therefore, the frequencies $\omega_k(E_k(t))$. Eq. (S13), when combined with Eq. (S9), describes the system evolution in the locked state on the long timescale as it is ringing down, neglecting any dynamics near the locked state.

Understanding the exact fully-coupled dynamics on the short timescale near the locked state and the general conditions for the existence of the persistent locked state is beyond the scope of this study. However, the PTS-like locked state for $\frac{d\omega_2}{dE_2} \neq 0$ can be semi-quantitatively understood in the simplified limit, Mode 2 is still treated as the nearly-constant external drive for Mode 1, but now with Mode 2 frequency varying with changes in its energy as a function of time. Similar to previously considering Mode 1 in the rotating frame of a constant frequency $\omega_2/n$, we now consider Mode 1 in the reference frame rotated at the variable frequency $\omega_2(E_2)/n$ defined by Mode 2. Neglect energy loss due to dissipation, $E_1 + E_2 = E_{1,res} + E_{2,res}$, and using Eq (S10) near a common resonance frequency we obtain

$$\omega_1(E_1) - \frac{\omega_2(E_2)}{n} = (E_1 - E_{1,res}) \frac{d\omega_1}{dE_1} - \frac{1}{n}(E_2 - E_{2,res}) \frac{d\omega_2}{dE_2} =$$
$$= (E_1 - E_{1,res}) \left(\frac{d\omega_1}{dE_1} + \frac{1}{n}\frac{d\omega_2}{dE_2}\right). \tag{S14}$$

Eq. (S14) has the same form as Eq. (S12) for the locked state dynamics with linear Mode 2, except for having an *effective* Mode 1 frequency-energy dependence

$$\left(\frac{d\omega_1}{dE_1}\right)_{effective} = \frac{d\omega_1}{dE_1} + \frac{1}{n}\frac{d\omega_2}{dE_2}. \tag{S15}$$

Therefore, Mode 1 dynamics near the locked state with both modes nonlinear may be quantitatively described similarly to the case with linear Mode 2, but using an effective nonlinearity for Mode 1.. However, we admit that a more consistent treatment of the dissipation for each mode, as well as a consistent fully coupled treatment of the system in general, has to be applied for a more accurate description. The analysis provided in this section is only intended to illustrate how the two-mode PTS-like locked state model we developed to describe our experiments might be generalized to include higher integer frequency ratio $n$ as well as an arbitrary eigenfrequency-energy dependencies $\omega_k(E_k)$ for each of the two nonlinear modes. In conjunction with Supplemental Section 4, this illustrates how, depending on each Mode's nonlinearities, the persistent locked state leads to either an increase or a decrease in the apparent observed decay rate of a given mode.

When appropriate nonlinear coupling terms exist, multiple modes and applied drive stimuli can resonantly exchange energy when the frequency matching conditions of the form $\sum m_k f_k = 0$ are satisfied for some integer $m_k$. Here the external stimuli are mathematically equivalent to infinitely stiff and massive lossless modes. Since the lowest-order nonlinear terms not forbidden by symmetry are often dominant, the coupling is strongest for the lowest number of participating modes and smallest integers. Such coupled nonlinear systems are commonly encountered and studied, including photonic, nanomechanical, cavity optomechanical, RF and other domains, where nonlinear signal generation and transduction are of high interest. These systems are less constrained and general conditions leading to the persistent locked states are unclear. However, the existence of such stable resonances is apparent, considering, for example, the case of Mode 1 being nonlinearly driven by a combination of Modes 2 and 3 subject to $nf_1 = f_2 + f_3$ and having high energy, stiffness and modal mass, such that the backaction of Mode 1 on them is negligible. Perhaps the simplest possible example of such interaction is encountered by subjecting the Mode 2 of our system to a constant linear drive. Initializing Mode 1 with sufficient energy to lock into a PTS leads to a system locked in a period-tripled symmetry-broken steady state.

## S8- The fully-coupled model of two nonlinearly coupled oscillators at the 1:3 internal resonance and how it is reduced to a PTS-like model

The free-ringdown system of two-coupled Duffing oscillators at the 1:3 internal resonance can be described by two equations of motion as [3,6]:

$$\ddot{q}_1 + \Gamma_1 \dot{q}_1 + \omega_{1,linear}^2 q_1 + \alpha_1 q_1^3 + m_1^{-1} \partial H_{int}/\partial q_1 = 0 \quad (S16x)$$
$$\ddot{q}_2 + \Gamma_2 \dot{q}_2 + \omega_{2,linear}^2 q_2 + \alpha_2 q_2^3 + m_2^{-1} \partial H_{int}/\partial q_2 = 0 \quad (S17x)$$

with an interaction energy term $H_{int} = g q_1^3 q_2$ for 1:3 internal resonance and the values of coupling constant $g$, modal masses $m_1$ and $m_2$ and coefficients $\alpha_1$ and $\alpha_2$ dependent on the choice of the scales of the generalized coordinates $q_1$ and $q_2$. Takin the experimentally measured voltage values directly as the generalized coordinates, the Duffing coefficients of the two modes obtained from their measured Duffing spectrum are $\alpha_1 \approx 2.41 \cdot 10^8 \text{ s}^{-2}\text{V}^{-2}$ and $\alpha_2 \approx -1.03 \cdot 10^7 \text{s}^{-2}\text{V}^{-2}$.

By taking the derivatives and defining coupling coefficients $g_{12} = \frac{g}{m_1}$ and $g_{21} = \frac{g}{m_2}$, we can write the equations of motion as:

$$\ddot{q}_1 + \Gamma_1 \dot{q}_1 + \omega_{1,linear}^2 q_1 + \alpha_1 q_1^3 + 3g_{12} q_1^2 q_2 = 0 \quad (S16)$$
$$\ddot{q}_2 + \Gamma_2 \dot{q}_2 + \omega_{2,linear}^2 q_2 + \alpha_2 q_2^3 + g_{21} q_1^3 = 0 \quad (S17)$$

when no external forces are applied.

It is instructive to consider the system in the rotating wave approximation $q_1 = u_1 e^{\frac{i\omega_{2,linear}t}{3}} + u_1^* e^{-\frac{i\omega_{2,linear}t}{3}}$ and $q_2 = u_2 e^{i\omega_{2,linear}t} + u_2^* e^{-i\omega_{2,linear}t}$, and neglecting non-resonant rapidly varying terms:

$$\dot{u}_1 + \frac{\Gamma_1}{2}u_1 + i\left(\frac{\omega_{2,linear}}{3} - \omega_{1,linear}\right)u_1 - i\frac{3\alpha_{11}}{2\omega_{1,linear}}|u_1|^2 u_1 - i\frac{3g_{12}}{\omega_{1,linear}}u_1^{*2}u_2 = 0 \quad (S18)$$

$$\dot{u}_2 + \frac{\Gamma_2}{2}u_2 - i\frac{3\alpha_{22}}{2\omega_{2,linear}}|u_2|^2 u_2 - i\frac{g_{21}}{\omega_{2,linear}}u_1^3 = 0 \quad (S19)$$

With the separately measured parameters, $\Gamma_1$, $\Gamma_2$, $\omega_{1,linear}$, $\omega_{2,linear}$, $\alpha_1$ and $\alpha_2$, we can fit the results using only two fitting parameters $g_{12}$ and $g_{21}$. For example, data in Fig. 2a can be fit by this complete model shown as Fig. S6-1 where $g_{12} \approx 3.32 \cdot 10^5 \text{s}^{-2}\text{V}^{-2}$, and $g_{21} \approx 2.98 \cdot 10^6 \text{s}^{-2}\text{V}^{-2}$, are fitted by achieving the least-residual between the fit and the amplitude data.

For comparing with experiment, we have chosen to use the experimental voltages directly as the generalized coordinates. This choice results in the unequal modal masses $m_1$ and $m_2$ for the two modes, and therefore unequal values of $g_{12} = \frac{g}{m_1}$ and $g_{21} = \frac{g}{m_2}$. For theoretical analysis, it is conventional to choose the generalized coordinates $\tilde{q}_1$ and $\tilde{q}_2$ such that $\tilde{m}_1 = \tilde{m}_2$ and therefore $\tilde{g}_{12} = \tilde{g}_{21}$. This can be easily done by rescaling the generalized coordinate for mode 2 as $\tilde{q}_2 = q_2\sqrt{\frac{m_2}{m_1}} = q_2\sqrt{\frac{g_{12}}{g_{21}}} \approx 0.33 q_2$, obtaining

$$\ddot{\tilde{q}}_1 + \Gamma_1 \dot{\tilde{q}}_1 + \omega_{1,linear}^2 \tilde{q}_1 + \alpha_1 \tilde{q}_1^3 + 3\tilde{g}\tilde{q}_1^2 \tilde{q}_2 = 0 \quad (S16*)$$
$$\ddot{\tilde{q}}_2 + \Gamma_2 \dot{\tilde{q}}_2 + \omega_{2,linear}^2 \tilde{q}_2 + \tilde{\alpha}_2 \tilde{q}_2^3 + \tilde{g}\tilde{q}_1^3 = 0 \quad (S17*)$$

with the rescaled Duffing coefficient $\tilde{\alpha}_2 = \alpha_2 \frac{m_1}{m_2} \approx 9.0\alpha_2$ and a common rescaled coupling coefficient $\tilde{g} = \sqrt{g_{12}g_{21}}$. We note that the interaction energy $H_{int}$, kinetic energies $\frac{1}{2}m_1\dot{q}_1^2$ and $\frac{1}{2}m_2\dot{q}_2^2$ and potential energies $\frac{1}{2}m_1\omega_{1,linear}^2 q_1^2$ and $\frac{1}{2}m_2\omega_{2,linear}^2 q_2^2$ are the coordinate-scale-invariant physical quantities and remain unchanged. The energy and phase results presented in Fig. S6 are also invariant with respect to coordinate scaling.

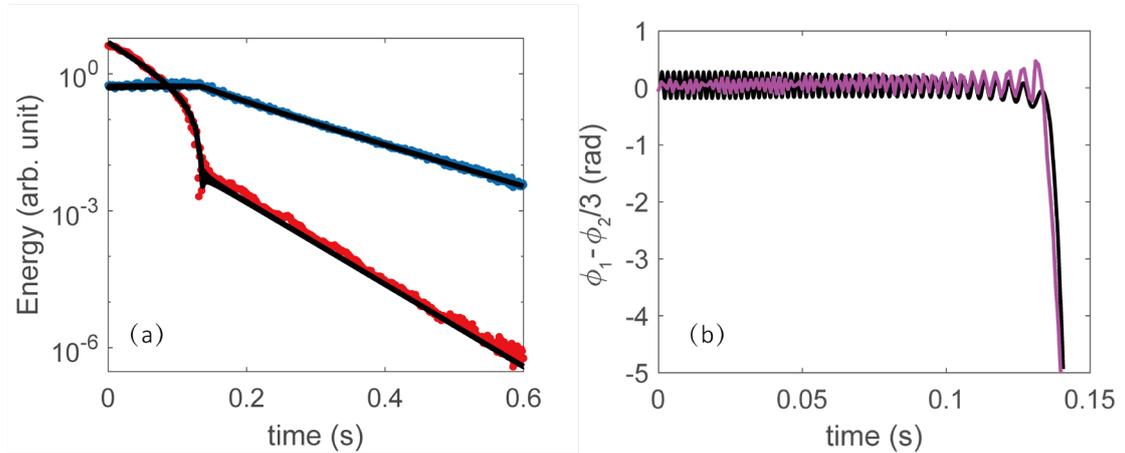

Fig. S6-1 Fit by complete model. The fitted results are shown as black lines for amplitude (a) and phase (b). The red and blue symbols and the purple line are the experimental data. The complete model, i.e. Eq. (18) and (19), agree with the measured energy data as well as the time-averaged energy evolution fit in Fig. 2a from the simple model. Beyond that, the complete model reproduces the experimentally observed fast phase oscillation as the mode1 rotates in the PTS pseudopotential with precession frequency $\Omega$, also shown in Movie 1.

Another option is using the calibration constant obtained from SI section 5 which is based on the assumption of constant energy flow $W$. The calibration constant in Eq. (S6) $\sqrt{c1/c2} = \frac{m_1}{m_2}$ is calibrated to be $\approx 8.8$, used to put the mode energies on the common energy scale in Fig. 2 and 3. By fixing the ratio $\frac{m_1}{m_2} = 8.8$, the two fitting parameters in Eq. (S16) and (S17) can be reduced to a single adjustable parameter $g$, where $g_{12} = g$ and $g_{21} = 8.8\ g$. Fitting it to the experiment results, as shown in Fig. S6-2, yields $g_{12} = g \approx 3.09 \times 10^5 \text{s}^{-2}\text{V}^{-2}$ and $g_{21} = 8.8g \approx 2.73 \times 10^6 \text{s}^{-2}\text{V}^{-2}$. Naturally, this single parameter fit gives nearly the same results as that of the complete two-fitting parameter fit ($g_{12} \approx 3.32 \cdot 10^5 \text{s}^{-2}\text{V}^{-2}$, and $g_{21} \approx 2.98 \cdot 10^6 \text{s}^{-2}\text{V}^{-2}$).

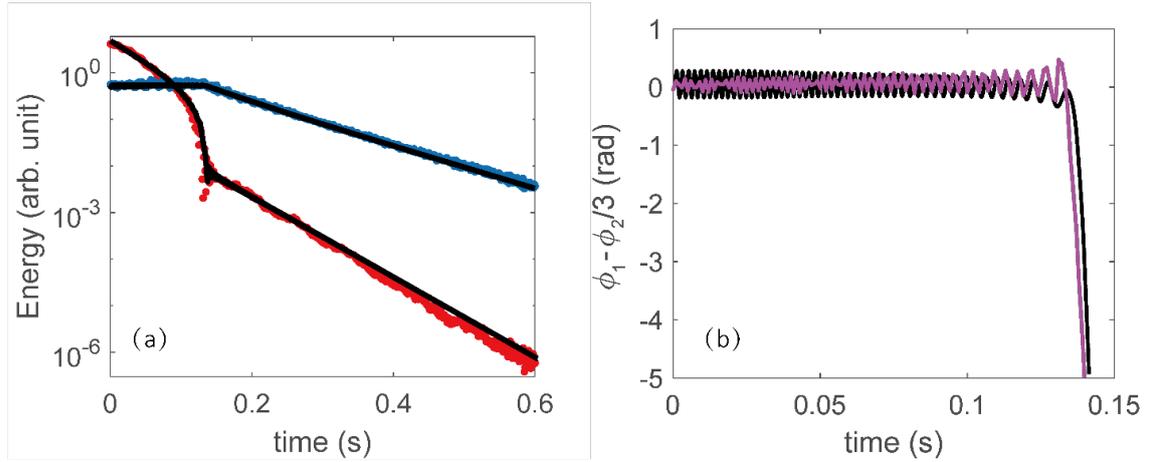

Fig S6-2. Modeling results with *a single coupling parameter* as the sole fit parameter, using the previously obtained calibration to a common energy scale[SI section 5]. The labels are the same as Fig. S6-1, the only difference is that we fix $g_{21} = 8.8g_{12}$ and fit a single parameter $g_{12}$. It generates $g_{12} \approx 3.09 \times 10^5 \text{s}^{-2}\text{V}^{-2}$ and $g_{21} = 8.8g_{12} \approx 2.73 \times 10^6 \text{s}^{-2}\text{V}^{-2}$.

Eq. S19 can be rewritten as

$$\dot{u}_2 + \left[\frac{\Gamma_2}{2} + \text{Im}\left(\frac{g_{21}}{\omega_{2,linear}} \frac{u_1^3}{u_2}\right)\right]u_2 - i\left[\frac{3\alpha_{22}}{2\omega_{2,linear}}|u_2|^2 + \text{Re}\left(\frac{g_{21}}{\omega_{2,linear}} \frac{u_1^3}{u_2}\right)\right]u_2 = 0 \quad \text{(S20)}$$

to explicitly express the interaction term as a change in the damping rate, $\text{Im}\left(\frac{g_{21}}{\omega_{2,linear}} \frac{u_1^3}{u_2}\right)$, and the frequency $\text{Re}\left(\frac{g_{21}}{\omega_{2,linear}} \frac{u_1^3}{u_2}\right)$.

Experimentally we observe that mode 2 frequency and amplitude change gradually, within the experimental resolution, in all regimes except very near the unlock time. This suggests considering Eq. S20 in the limit where the rate term $\frac{g_{21}}{\omega_{2,linear}} \frac{u_1^3}{u_2}$ is small. In this limit, mode 2 oscillates at the frequency dictated by its linear frequency and Duffing nonlinearity backbone, regardless of the dynamics of mode 1 (seen in Fig. 2d, 3b, red dots), i.e. mode 1 dynamics barely affect the frequency and phase of mode 2. Specifically, the phase of mode 2 oscillation becomes fully determined by the Duffing nonlinearity and changes slowly with the decreasing mode 2 amplitude, on the timescale determined by the damping rates of the system. This allows considering Eq 1 for mode 1 with interaction force of slowly changing amplitude and frequency, and applying the insights that have been previously developed for a single mode period tripling state [4].

Within this single-mode PTS description, near one of three stationary lock points, a high-Q mode (e.g. mode 1) undergoes precession in the local pseudopotential at a frequency $\Omega$, which defines the fastest dynamical timescale. The approximate PTS-like description is applicable when the force changes slowly on this timescale, i.e. both the PTS stationary point location and the pseudopotential profile change adiabatically. Qualitatively, the PTS description is self-consistent when the rate term $\frac{g_{21}}{\omega_{2,linear}} \frac{u_1^3}{u_2}$ describing modification to mode 2 dynamics by mode 1 is much smaller than $\Omega$.

While mode 2 frequency is barely affected by Mode 1, its energy loss rate to mode 1, $2\text{Im}\left(\frac{g_{21}}{\omega_{2,linear}} \frac{u_1^3}{u_2}\right)$, can be comparable to its weak intrinsic damping $\Gamma_2$. Therefore we retain this part of the interaction, which leads to the energy flow $W$ between modes in Eq. (S6), and increases the overall mode 2 loss rate, as seen in Figure 2a, red, and 3e, red dots.

To summarize, at the above analytically discussed and experimentally demonstrated limit, the major dynamics of mode 2 (including phase/amplitude) in the rotating frame are on the time scale of $\sim 1/\Gamma_{1,2}$. Given the low energy loss rate $1/\Gamma_{1,2}$, mode 2 acts as an adiabatically varying nonlinear drive to mode 1. Mode 1 dynamics can be described using the PTS picture with a slowly varying period-3 force at the frequency of $\omega_2 \approx 3\omega_1$. Specifically, from the perspective of mode 1 at this limit, mode 2 provides a period-3 harmonic drive with slowly varying amplitude and frequency, i.e. $q_2(t) \propto F_2(t)\cos[\phi_2(t)]$ where $\dot{\phi}_2(t) = \omega_2$. As the result, Eq. (S16) is reduced to Eq. (1) where the coupling term becomes $q_1^2 F_2(t)\cos[\phi_2(t)]$ from $3g_{12}q_1^2 q_2$.

The PTS model provides a valid description of our system, evident by the slowly evolving $F_2(t)$ and $\omega_2(t)$ (given by the measured Mode 2 dynamics) relative to the fast frequency and the small-amplitude oscillations of the locked Mode 1, and frequency-amplitude relationship defined by Duffing backbones. However, as shown by the numerical modeling using the fully-coupled model in Fig. S7, the locking phenomena can even occur with parameters where the PTS model does not strictly apply. A general approach, such as Ref. [6], can be used to explore the broader parameter range where the discovered phase-locking phenomenon can occur. Here we illustrate the generality of the observed phenomena by numerically simulating the fully-coupled model with coupling strength, Duffing nonlinearities, and damping rates varied within two orders of magnitude. As shown in Fig. S7, the phase-locking state remains there, showing this physical behavior is present across a wide range of multiple system parameters. In Fig. S7(b), (c), (e), (g), and (j), although the condition of PTS locking by a force with a smoothly varying amplitude and frequency already breaks down, evident by the oscillating Mode 2 amplitude, the phase-locked state still exists, evident by the sustained constant time-averaged Mode 1 amplitude and constant relative phase.

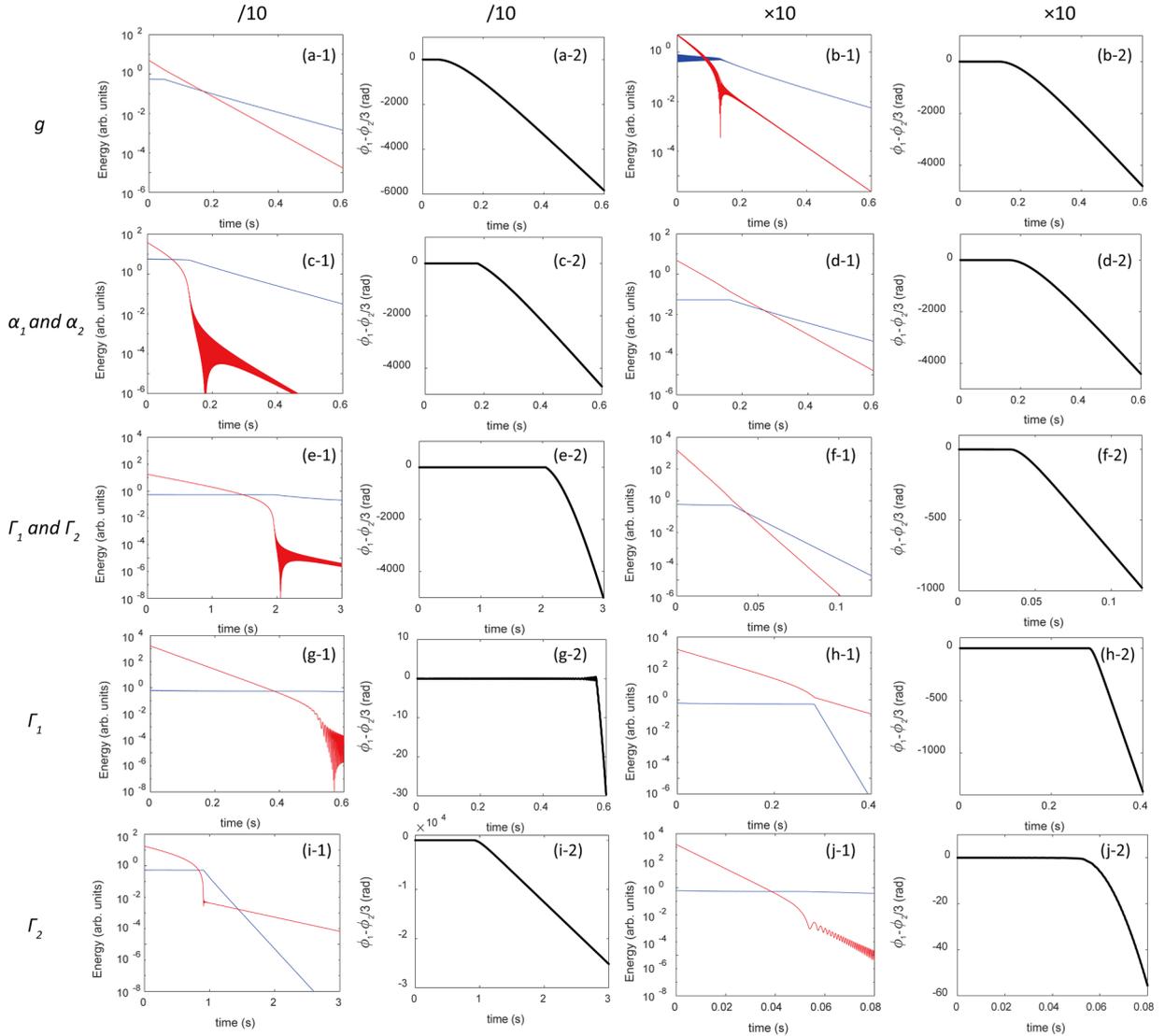

Figure S7. Simulated energy and relative phase of the two modes using the fully-coupled two-mode model with parameters scaled as follows relative to the ones used for Figure S6-2: $g/10$ (a); $g \times 10$ (b); $\alpha_1/10$ and $\alpha_2/10$ (c); $\alpha_1 \times 10$ and $\alpha_2 \times 10$ (d); $\Gamma_1/10$ and $\Gamma_2/10$ (e); $\Gamma_1 \times 10$ and $\Gamma_2 \times 10$ (f); $\Gamma_1/10$ (g); $\Gamma_1 \times 10$ (h); $\Gamma_2/10$ (i); $\Gamma_2 \times 10$ (j); Within a wide range of the parameters, the persistent locked states are shown. The blue and red lines denote modes 1 and 2's energy, respectively. The black line denotes their relative phase. When the system is initialized with $E_2 \gg E_1$ and the relative phase near the expected PTS locked state, an extended phase-locked period of time with near-constant Mode 1 amplitude is observed for each parameter set.

The description of our system motion can be further simplified from the complete model when an additional condition is valid, namely that Mode 1 also approximately follows its Duffing backbone, i.e. the modification of Mode 1 frequency-amplitude relation by the interaction term is small. This condition is valid in our system since the measured frequency-amplitude relation in Fig. 2 and Fig. 3 are consistent with its Duffing backbone. This condition means that the time-averaged interaction energy is negligible compared to either of the mode energies, including in the locked PTS state. As the result, the extra energy loss from Mode 2 in addition to natural damping equals the extra energy gain of Mode 1, as described by Eq. (S6). Meanwhile, the phase-locked mode amplitudes are fully determined by their common frequency at all times. The time evolution of the amplitudes is determined by the intrinsic energy loss rates decreasing the total energy of the system, while the energy flow between the modes is such that their individual amplitudes remain on their backbones at some common frequency changing in time. In particular, when the mode 2 backbone is linear (near-vertical), mode 1 time-averaged frequency must remain constant to maintain the lock, which results in mode 1 time-averaged amplitude remaining constant during lock as shown in Fig. 2a. Eq. (S6) with a single parameter W achieves good agreement for the energy-time evolution in Fig 2, Fig. 3, and the coherent time fit in Fig 4 (Eq. S8).

While this energy transfer view is particularly compelling in its simplicity, it is not a necessary condition for the PTS description of the Mode 1 evolution. The only benefit of this condition in our work is providing a simple and intuitive way to quantitatively fit the slow energy evolution of both modes and the coherent time, while neglecting the faster features of the full PTS model dynamics from Mode 1, such as the small but observable oscillations of Mode 1 phase and amplitude at the precession frequency $\Omega$.

In summary, while we have numerically confirmed that the system can be modeled well with fully-coupled equations, the reduced model advances the physical understanding of the mode interaction process, while providing a good quantitative description. The reduced model makes a connection between fully-coupled modes and the PTS for the lower mode, which intuitively explains the novel dynamics observed in this work: initial-phase-dependent bypassing or phase-locking at $2\pi/3$ relative phase increment; frequency and amplitude evolution, including the non-monotonic lower mode energy evolution; long coherent time and locking probability as a function of initial conditions.

## S9- Dynamics of Mode 1 in the PTS pseudopotential and initial-phase-dependent trajectory

The dynamics of Mode 1 in the PTS pseudopotentials are shown in Fig. S8. Fig. S8(a) presents in the phase diagram, four groups of experimental data obtained in the locking/bypassing experiment (Fig. 3). Three groups lock to the PTS Basins around distinct stationary points, while one group (black) bypasses the PTS and directly decays to the ground state. Due to the strong Duffing nonlinearity, there is a rapid frequency change with amplitude, squeezing the Basin in the radial direction. Therefore, we re-plot the same data in Fig. S8(b), zooming in on the data at $R > 0.033$ (labeled by the red dashed line in panel (a)), by redefining the radial coordinate as $Q' + iP' = (R -$

$0.033)e^{i\varphi}$. The trajectory of ringdown is consistent with our schematic figure in Fig. 1(e), clearly demonstrating the locking/bypassing trajectory in the PTS pseudopotentials, and the precession($\Omega$) and decay($\kappa$) of Mode 1, as is also seen in Movie 1.

To illustrate the initial phase dependence of the phase-locking/bypassing, in Figure S8(c), we present the 44 series of data from all those experiments where locking was observed out of the 100 total repeat experiments starting from random, uniformly distributed phases. The conditions correspond to the $A_{2,0}$ = 0.03 data in Fig. 4(b) [the set with locking probability ≈ 50 %]. We focus on the data just before entering the basin regions, by further zooming in and setting origin at $R$ = 0.040, i.e. $Q'' + iP'' = (R - 0.040)e^{i\varphi}$, and truncating the data traces to include only *R>0.042* (red dashed line in (b)). The 44 groups of locking data are colored based on which Basin they end up with. The remaining 56 groups of bypassing data, not shown here, concentrate within the white spaces between the discrete colored spirals, uniformly filling the phase space due to the uniform distribution of the initial relative phase.

Only initial conditions within the basin of attraction for each PTS stationary point (areas covered by the locking data points) lead to a locked state while other initial conditions (white empty areas) result in bypassing/ground-state, consistent with the illustration in Fig. 1(e) and Fig. 4(c) (d). In the paper, we use the schematic generated with smaller frequency detuning to more clearly illustrate all the features on a single radial scale (no need to zoom in) for facilitating understanding.

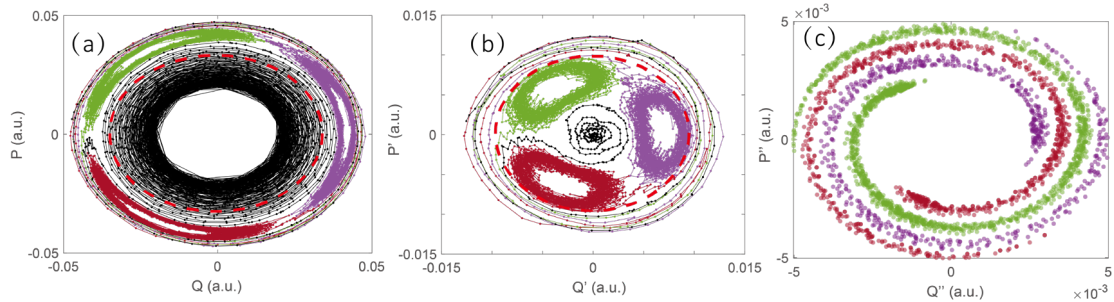

Figure S8. The dynamics of Mode 1 in the PTS pseudopotential with a reference phase of $\phi_2(t)/3$. (a) Four groups of locking/bypassing data taken at the same experimental conditions as shown in Fig. 3. (b) Same as (a) but rescaling the radial scale by setting origin at $R$ = 0.033 (red dashed circle in (a)) instead of 0. Three groups of data lock and precess around three PTS stationary points, while one group bypasses and rings down to the ground state. (c) *44 groups* of experimental data that resulted in locking out of 100 repeated experiments with uniformly-distributed random initial phases. They are colored based on which basin they end up with. The other 56 groups of data (not shown) are all bypassing the lock. The bypassing data concentrate within the white space between the colored spirals. It is comparable to Fig. 4d. For (c), we focus on the data just before entering the basin regions, by further zooming in and setting origin at $R$ = 0.040, i.e. $Q'' + iP'' = (R - 0.040)e^{i\varphi}$, and truncating the data traces to include only *R>0.042* (red dashed line in (b)).

## S10- Mechanical spectrum of the device

We drive and measure the mechanical spectrum of the resonator up to 1 MHz (≈15x the fundamental frequency) shown as the blue dots in the upper panel of Fig. S9. The gray dots are the measurement noise floor measured at the stationary reference substrate. Besides, we numerically simulate the eigenfrequency of the devices and quantitatively match all the modeled mechanical modes of the system, shown as the lower panel of Fig. S9, eliminating the possibility of hidden modes. The dashed line labels $n \times \omega_{IR}$ where $\omega_{IR} = \omega_{2,linear}/3$ is the internal resonance frequency and $n = 1,2,3,...$ Except for mode 1 and mode 3 (labeled red), there is no other commensurate frequency matching. As described on the paper Mode 1's linear eigenfrequency is detuned lower than $\omega_{IR}$, but it matches $\omega_{IR}$ at a specific amplitude due to its Duffing-nonlinearity frequency shift.

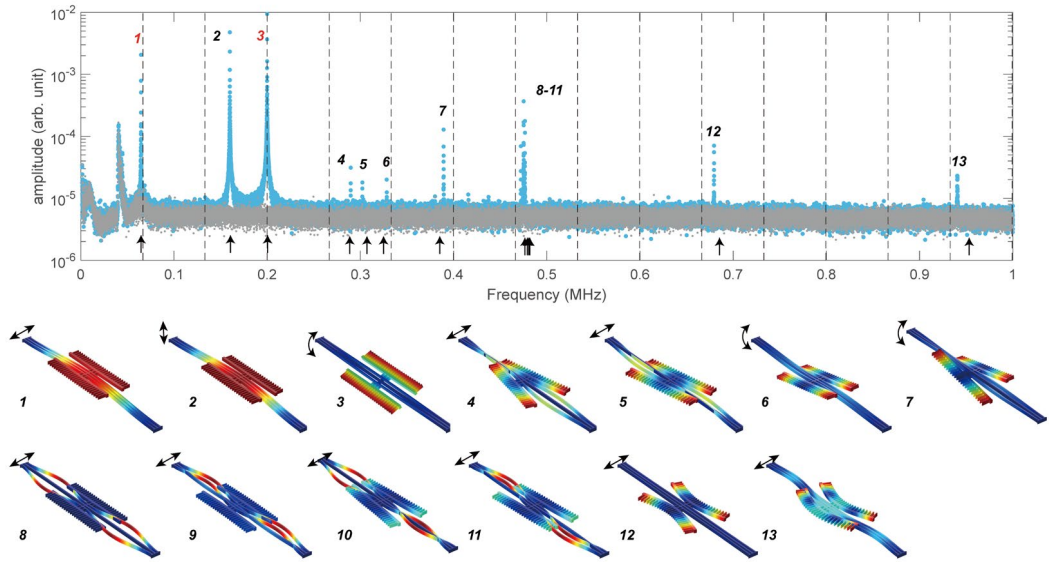

Figure S9. Measured mechanical spectrum (blue dots) and simulated mode shape are shown as upper and lower panels. The simulated (up-pointing black arrows) and measured eigenfrequencies match each other, with disagreement <5 % due to fabrication imperfections. Dashed lines are integer multiples of the internal resonance frequency. They indicate there is no other matching frequency at the 1:3 internal resonance. The gray dots are the measurement noise floor. The arrows in the modal shape panel depict in-plane flexural modes, out-of-plane flexural modes, and torsional modes. Mode 8 to 11 are four degenerate beam modes.

# References

[1] D. Antonio et al. "Frequency stabilization in nonlinear micromechanical oscillators" Nat Commun **3**, 806 (2012).

[2] C. Chen et al. "Direct observation of coherent energy transfer in

nonlinear micromechanical oscillator" Nat Commun **8**, 15523 (2017).

**[3]** J. Guttinger et al. "Energy dependent path of dissipation in nanomechanical resonators" Nature Nanotech **12**, 631-636 (2017).

**[4]** J. Gosner et. al. "Relaxation dynamics and dissipative phase transition in quantum oscillators with period tripling" Phys. Rev. B **101**, 054501 (2020).

**[5]** Y. Zhang et al. "Time-translation-symmetry breaking in a driven oscillator: From the quantum coherent to the incoherent regime" Phys. Rev. A **96**, 052124 (2017)

**[6]** O. Shoshani et al. "Anomalous decay of nanomechanical modes going through nonlinear resonance" Sci Rep **7**, 18091 (2017)